\documentclass[aps,prd,showkeys,preprint,amsmath,amssymb,nofootinbib]{revtex4-1}

\usepackage{graphicx}
\usepackage{comment}
\usepackage{bm} 
\usepackage{slashed}
\usepackage{marvosym}
\usepackage{textcomp}
\usepackage{threeparttable}
\usepackage{booktabs}
\usepackage{setspace}
\usepackage{latexsym}
\usepackage{amsmath,amsthm,amssymb}
\usepackage{color}
\usepackage[colorlinks=true, pdfstartview=FitV, linkcolor=red, citecolor=blue, urlcolor=blue, pdftitle={text},pdfauthor={Yoshimasa Hidaka},pdfsubject={}, pdfkeywords={}]{hyperref}
\usepackage{scalerel,stackengine}
\usepackage[normalem]{ulem} 
\usepackage[margin=20pt,small]{caption}

\allowdisplaybreaks[1]
\stackMath
\newcommand\reallywidehat[1]{%
	\savestack{\tmpbox}{\stretchto{%
			\scaleto{%
				\scalerel*[\widthof{\ensuremath{#1}}]{\kern-.6pt\bigwedge\kern-.6pt}%
				{\rule[-\textheight/2]{1ex}{\textheight}}
			}{\textheight}%
		}{0.5ex}}%
	\stackon[1pt]{#1}{\tmpbox}%
}
\renewcommand{\TPTtagStyle}%
{\normalsize\textit}
\newcommand{\ri}{\mathrm{\rm i}}

\newcommand{\arXiv}[1]{\href{http://arxiv.org/abs/#1}}

\newcommand{\tr}{\mathrm{tr}\,}

\newcommand{\sgn}{\mathop{\mathrm{sgn}}}

\newcommand{\nn}{\nonumber}
\newcommand{\gam}{\gamma}
\renewcommand{\a}{\alpha}
\renewcommand{\b}{\beta}

\newcommand{\selfEnergy}{\Sigma}

\begin{document}
\title{Wigner functions and quantum kinetic theory of polarized photons}
\author{Koichi Hattori$^{1}$, Yoshimasa Hidaka$^{2,3,4}$, Naoki Yamamoto$^{5}$, Di-Lun Yang$^{5}$}
\affiliation{
$^1$Yukawa Institute for Theoretical Physics, Kyoto University, Kyoto 606-8502, Japan.\\
$^2$KEK Theory Center, Tsukuba 305-0801, Japan.\\
$^3$Graduate University for Advanced Studies (Sokendai), Tsukuba 305-0801, Japan.
\\$^4$RIKEN iTHEMS, RIKEN, Wako, Saitama 351-0198, Japan.
\\$^5$Department of Physics, Keio University, Yokohama 223-8522, Japan.\\} 

\preprint{YITP-20-129, KEK-TH-2262, J-PARC-TH-0228}
\begin{abstract}
We derive the Wigner functions of polarized photons in the Coulomb gauge with the $\hbar$ expansion applied to quantum field theory, 
and identify side-jump effects for massless photons. 
We also discuss the photonic chiral vortical effect for the Chern-Simons current and zilch vortical effect for the zilch current in local thermal equilibrium as a consistency check for our formalism. The results are found to be in agreement with those obtained from different approaches.
Moreover, using the real-time formalism, we construct the quantum kinetic theory (QKT) for polarized photons. By further adopting a specific power counting scheme for the distribution functions, we provide a more succinct form of an effective QKT. This photonic QKT involves quantum corrections associated with self-energy gradients in the collision term, which are analogous to the side-jump corrections pertinent to spin-orbit interactions in the chiral kinetic theory for massless fermions. The same theoretical framework can also be directly applied to weakly coupled gluons in the absence of background color fields.             
\end{abstract}

\maketitle
\section{Introduction}
Quantum transport of circularly polarized photons is a fundamental issue in various areas of physics from optics, photonics, condensed matter physics, and nuclear physics to astrophysics.
One well-known example is the spin-dependent deflection of photons due to the spin-orbit interaction, called the photonic spin Hall effect \cite{Bliokh:2004gz,Onoda:2004zz}. 
Another recently found example is the photonic helicity current induced by vorticity, called the photonic chiral vortical effect (CVE) \cite{Avkhadiev:2017fxj,Yamamoto:2017uul,Zyuzin:2017aa,Huang:2018aly,Prokhorov:2020okl}. However, the photonic CVE defined through the Chern-Simons (CS) current is not locally gauge-invariant. One can instead consider its gauge-invariant version called the zilch vortical effect (ZVE) \cite{Chernodub:2018era,Copetti:2018mxw} by making use of the so-called zilch \cite{Lipkin_Zilch,morgan1964two,Kibble_Zilch} as an infinite set of conserved quantities in non-interacting Maxwell's theory.%
\footnote{See also Ref.~\cite{PhysRevLett.104.163901}, where it is shown that the lowest-order zilch characterizes optical chirality for photons interacting with chiral molecules.}

The conventional approach to describe the quantum transport of photons out of equilibrium is based on the semi-classical equation of motion including the effects of the Berry phase \cite{Bliokh:2004gz,Onoda:2004zz,Yamamoto:2017uul,Huang:2018aly}. In equilibrium, one may alternatively compute, e.g., the photonic helicity current from quantum field theory \cite{Avkhadiev:2017fxj,Chernodub:2018era,Copetti:2018mxw,Prokhorov:2020okl}.
To the best of our knowledge, however, the generic quantum kinetic theory (QKT) for non-equilibrium many-body photons with collisional effects has not been well established based on the underlying quantum field theory---quantum electrodynamics (QED). 

In the case of Weyl or Dirac fermions, there have been substantial progresses in the developments of relativistic QKT for massless \cite{Son:2012wh,Stephanov:2012ki,Son:2012zy,Chen:2012ca,Manuel:2013zaa,Chen:2014cla,Chen:2015gta,Hidaka:2016yjf,Hidaka:2017auj,Mueller:2017lzw,Mueller:2017arw,Huang:2018wdl,Carignano:2018gqt,Dayi:2018xdy,Liu:2018xip,Lin:2019ytz,Carignano:2019zsh} and massive cases \cite{Mueller:2019gjj,Weickgenannt:2019dks,Gao:2019znl,Hattori:2019ahi,Wang:2019moi} with applications to quark-gluon plasmas (QGP) in heavy ion collisions \cite{Kharzeev:2016sut,Huang:2017tsq,Sun:2017xhx,Hidaka:2018ekt,Sun:2018idn,Yang:2018lew,Liu:2019krs,Shi:2020htn}, Weyl semimetals \cite{Son2013a,Basar:2013iaa,Landsteiner:2013sja,Gorbar:2016ygi}, core-collapse supernovae \cite{Yamamoto:2020zrs}, and cosmology \cite{Yamamoto:2020phl}. For example, it was shown that the CVE of Weyl fermions \cite{Vilenkin:1979ui,Erdmenger2009,Banerjee:2008th,Son:2009tf,Landsteiner:2011cp} is in connection to the Berry phase and the so-called side-jump phenomenon \cite{Chen:2014cla,Chen:2015gta,Hidaka:2016yjf,Hidaka:2017auj}. Such a connection is revealed in the Wigner-function approach based on quantum field theory along with the $\hbar$ expansion \cite{Hidaka:2016yjf,Hattori:2019ahi}. 

It is thus tempting to explore a similar scenario for polarized photons. In Ref.~\cite{Huang:2020kik}, the Wigner functions of polarized photons and corresponding CVE and ZVE have been recently investigated. However, the Wigner functions were derived from the mixture of right and left-handed polarized photons via Maxwell's equations and gauge constraints therein. It is still desirable to obtain the Wigner functions constructed individually from right/left-handed polarized photons through a first-principle derivation from QED to make a direct comparison with fermionic Wigner functions in Weyl bases studied in Ref.~\cite{Hidaka:2016yjf}. Moreover, a corresponding QKT for tracking spin transport of polarized photons with collisions is also needed. 

In this paper, by exploiting the spinor-helicity formalism to write the covariant form of polarization vectors for right/left-handed photons \cite{Kleiss:1985yh,Gunion:1985vca,Xu:1986xb,Peskin:2011in}, we explicitly derive the Wigner functions up to $\mathcal{O}(\hbar)$ in the Coulomb gauge, which manifest how the Berry connections are encoded in distribution functions as the case for Weyl fermions. 
We also compute the photonic CVE and ZVE in local thermal equilibrium and confirm that our results are consistent with the previous results in Refs.~\cite{Chernodub:2018era,Copetti:2018mxw,Huang:2020kik}.
By using the real-time formalism and adopting a specific power counting scheme, we further construct the general form of the effective QKT for polarized photons with the collision term characterized by self-energies, similar to the fermionic case in Ref.~\cite{Yang:2020hri}. 

This formalism can also be directly applied to weakly coupled gluons in the absence of background color fields. This would pave the way to future study of entangled spin transport of quarks and gluons in QGP. Motivated by recent experimental observations of global polarization of $\Lambda$ hyperons in heavy ion collisions \cite{STAR:2017ckg,Adam:2018ivw,Abelev:2007zk1}, this direction should be important to understand how the dynamical evolution of the quark spin will be converted to the local spin polarization of hadrons \cite{Zhang:2019xya,Li:2019qkf,Kapusta:2019sad,Yang:2020hri,Weickgenannt:2020aaf,Hou:2020mqp,Bhadury:2020cop,Wang:2020pej} along the direction of the strong vorticity generated in peripheral collisions; see other theoretical works on developments of hydrodynamics with spin \cite{Florkowski:2017ruc,Montenegro:2017rbu,Florkowski:2018fap,Hattori:2019lfp,Fukushima:2020ucl} and statistical quantum field theory \cite{Becattini2013a, Becattini:2018duy, Becattini:2020qol}, which also aim at exploring underlying mechanisms and reconciling the existing tension between theoretical predictions and experimental observations for local spin polarization in heavy ion collisions (see Ref.~\cite{Becattini:2020ngo} for a recent review and more references therein).

The paper is organized as follows. In Sec.~\ref{sec_spinor_helicity}, we briefly review the spinor-helicity formalism and introduce the polarization vectors in the Coulomb gauge. In Sec.~\ref{sec_WF_photons}, we accordingly derive the Wigner functions for photons up to $\mathcal{O}(\hbar)$. 
As a check on the formalism that we develop in Sec.~\ref{sec_WF_photons}, in Sec.~\ref{sec_CS_current} we analyze the CVE/ZVE for photons in local equilibrium, confirming previous results from the literature. 
In Sec.~\ref{sec_QKT_photons}, we derive the QKT for photons and its effective version with specific power counting. Finally, we make short summary and outlook in Sec.~\ref{sec_sum_outlook}.     

Throughout the present paper, we use the mostly minus signature of the Minkowski metric $\eta^{\mu\nu} = {\rm diag} (1, -1,-1,-1)$, 
the completely antisymmetric tensor $ \epsilon^{\mu\nu\rho\lambda}$ with $ \epsilon^{0123} = 1$, 
and $\gamma_5={\rm i}\gamma^0\gamma^1\gamma^2\gamma^3$. 
We also use the notations $A^{(\mu}B^{\nu)}\equiv A^{\mu}B^{\nu}+A^{\nu}B^{\mu}$ 
and $A^{[\mu}B^{\nu]}\equiv A^{\mu}B^{\nu}-A^{\nu}B^{\mu}$.

\section{Spinor-helicity formalism for photons}\label{sec_spinor_helicity}

Let us first briefly recapitulate essential parts of the so-called spinor-helicity formalism~\cite{Kleiss:1985yh,Gunion:1985vca,Xu:1986xb} 
(see, e.g., Ref.~\cite{Peskin:2011in} for a review) that will be used in our following computations.
The basic idea of this formalism is to express spin-one vector fields as bispinors 
since they transform in the $(1/2, 1/2)$ representation of the Lorentz group. 
As an advantage of this formalism, we can avoid the redundancy to embed a massless photon 
with two physical degrees of freedom into a four-component vector field $A^{\mu}(x)$.
As we will show below, this formalism naturally allows us to obtain the result of 
the quantum kinetic theory for spin-one photons in the same form as that for fermions.

In this formalism, the polarization vectors of photons are written 
with fermion spinors as~\cite{Kleiss:1985yh,Gunion:1985vca,Xu:1986xb} 
\begin{eqnarray}
\label{spinor}
\epsilon^{\rm R}_{\mu}(p)=\frac{1}{\sqrt{4k\cdot p}}\bar{u}_{\rm R}(k)\gamma_{\mu}u_{\rm R}(p),\qquad
\epsilon^{\rm L}_{\mu}(p)=\frac{1}{\sqrt{4k\cdot p}}\bar{u}_{\rm L}(k)\gamma_{\mu}u_{\rm L}(p)\,,
\end{eqnarray} 
for the right-handed and left-handed helicity, respectively. 
Here, $p$ is the momentum of a photon, while $k$ is an auxiliary light-like vector such that $p\cdot k\neq 0$ and $p^2=k^2=0$.%
\footnote{We adopt the same convention as in Ref.~\cite{Peskin:2011in}.} 
The helicity eigenstates of massless fermions satisfy $\slashed{p}(1+\gamma_5)u_{\rm R}(p)/2=(p_0-|{\bm p}|)u_{\rm R}(p)$ and $\slashed{p}(1-\gamma_5)u_{\rm L}(p)/2=(p_0-|{\bm p}|)u_{\rm L}(p)$ in the Weyl representation. In general, one may replace $p_0$ by $n\cdot p$ and $\bm p$ by the component transverse to $n^{\mu}$ 
which is a timelike vector specifying a Lorentz frame for the spin basis. 
Here we simply choose $n^{\mu}=(1,\bm 0)$.
By using 
\begin{eqnarray}
u_{\rm R}(p)\bar{u}_{\rm R}(p)+u_{\rm L}(p)\bar{u}_{\rm L}(p)=\slashed{p}
\end{eqnarray}
for on-shell photons, one can show 
\begin{eqnarray}\label{Gmunu_with_k}
\epsilon_{\rm R}^{\mu}\epsilon_{\rm R}^{\nu*}+\epsilon_{\rm L}^{\mu}\epsilon_{\rm L}^{\nu*}=-\eta^{\mu\nu}+\frac{p^{\mu}k^{\nu}+k^{\mu}p^{\nu}}{p\cdot k}\,.
\end{eqnarray}
We may now assign a proper $k^{\mu}$ that meets the gauge choice. 
For example, we can take the Coulomb gauge $\partial_{\perp\alpha}A^{\alpha}=0$, where a transverse projection 
is defined as $v_{\perp}^{\mu}\equiv(\eta^{\mu\nu}-n^{\mu}n^{\nu})v_{\nu}$ for an arbitrary vector $v^{\mu}$.%
\footnote{More generically, one may take a gauge condition $\partial_{\mu} (\eta^{\mu \nu} - \ell^{\mu} \ell^{\nu}) A_{\nu}=0$, 
where $\ell^{\mu}$ is a unit timelike vector ($\ell^2=1$) that is not necessarily equal to $n^{\mu}$.
For convenience, we will choose ${\ell}^{\mu} = n^{\mu}$ throughout the paper.}
We also use shorthand notations $|{\bm v}|\equiv\sqrt{|v_{\perp}^2|}$ and $\hat{v}_{\perp\mu}\equiv v_{\perp\mu}/|{\bm v}|$. 
Then, the polarization sum is supposed to have the form 
\begin{eqnarray}
\epsilon^{\rm R}_{\mu}\epsilon^{{\rm R}*}_{\nu}+\epsilon^{\rm L}_{\mu}\epsilon^{{\rm L}*}_{\nu}
=-\eta_{\mu\nu}-\frac{p_{\perp\mu}p_{\perp\nu}}{|{\bm p}|^2}+\frac{(p\cdot n)^2 n_{\mu} n_{\nu}}{|{\bm p}|^2}
=-\eta_{\mu\nu}+\frac{1}{|{\bm p}|^2}p_{(\mu}\Big(p\cdot n n_{\nu)}-\frac{p_{\nu)}}{2}\Big)\,.
\end{eqnarray}
Comparing this expression with Eq.~(\ref{Gmunu_with_k}), one should thus take
\begin{eqnarray}
\frac{k_{\nu}}{k\cdot p}=\frac{1}{|{\bm p}|^2}\Big(p\cdot n n_{\nu}-\frac{p_{\nu}}{2}\Big)=\frac{1}{2|{\bm p}|}\big(n_{\nu}-\hat{p}_{\perp\nu}\big)\,,
\end{eqnarray}
which implies
\begin{eqnarray}\label{k_in_Coulomb}
k_{\nu}=k\cdot n \big(n_{\nu}-\hat{p}_{\perp\nu}\big)
\, .
\end{eqnarray}

In terms of the two-component spinors $c_{\rm R, L}(p)$, defined respectively as the lower and upper two-component
fields of $u_{\rm R, L}(p)/\sqrt{2|{\bm p}|}$, we have 
\begin{eqnarray}
\label{epsilon_in_c}
\epsilon_{\mu}^{\rm R}(p)=\frac{\sqrt{|{\bm p}||{\bm k}|}}{\sqrt{k\cdot p}}c^{\dagger}_{\rm R}(k)\sigma_{\mu}c_{\rm R}(p)\,,\qquad
\epsilon_{\mu}^{\rm L}(p)=\frac{\sqrt{|{\bm p}||{\bm k}|}}{\sqrt{k\cdot p}}c^{\dagger}_{\rm L}(k)\bar{\sigma}_{\mu}c_{\rm L}(p)\,.
\end{eqnarray}
Let us focus on $\epsilon^{{\rm R}}_{\mu}$. 
In the frame $n^{\mu}=(1,{\bm 0})$, the explicit forms of $c_{\rm R}(p)$ and $c_{\rm R}(k)$ are (see, e.g., Ref.~\cite{Hidaka:2016yjf})
\begin{eqnarray}
c_{\rm R}(p) = 
\begin{pmatrix}
\sqrt{\frac{|{\bm p}|+p^3}{2|{\bm p}|}} 
\\
\frac{p^1+{\rm i}p^2}{\sqrt{2|{\bm p}|(|{\bm p}|+p^3)}} 
\end{pmatrix}\,, \qquad
c_{\rm R}(k) = 
\begin{pmatrix}
\sqrt{\frac{|{\bm p}|-p^{3}}{2|{\bm p}|}}
\\
-\frac{p^1+{\rm i}p^2}{\sqrt{2|{\bm p}|(|{\bm p}|-p^3)}} 
\end{pmatrix},\qquad
\frac{\sqrt{|{\bm p}||{\bm k}|}}{\sqrt{k\cdot p}}=\frac{1}{\sqrt{2}}\,.
\end{eqnarray}
Then, one can show that
\begin{eqnarray}
c_{\rm R}(p)c_{\rm R}^{\dagger}(p) + c_{\rm R}(k)c_{\rm R}^{\dagger}(k)=I,
\qquad 
c^{\dagger}_{\rm R}(p)c_{\rm R}(k)=c^{\dagger}_{\rm R}(k)c_{\rm R}(p)=0,
\end{eqnarray}
where $I$ is a unit matrix. 
It turns out that $c_{\rm R}(k)$ corresponds to the eigenvector of right-handed fermions with negative energy.
Accordingly, we may denote $c_{\rm R}(p)=c_{\rm R}^{(+)}$ and $c_{\rm R}(k)=c_{\rm R}^{(-)}$ for convenience, where we use the indices ``$(\pm)$" to represent the eigenvectors of right-handed fermions with positive and negative energies, respectively. We thus arrive at the expression
\begin{eqnarray}
\label{epsilon^R}
\epsilon^{\rm R}_{\mu}(p)=\frac{1}{\sqrt{2}}c^{(-)\dagger}_{\rm R}(p)\sigma_{\mu}c^{(+)}_{\rm R}(p)\,.
\end{eqnarray}

\section{Wigner Functions for Polarized Photons}\label{sec_WF_photons}
Given the polarization vectors of right/left-handed photons, we are able to quantize the polarized gauge fields and 
compute the corresponding Wigner functions. 
Similar to the case for massless fermions, we can separate the right and left-handed sectors in the free theory. 
We start from the mode decomposition of a U(1) gauge field with the right- or left-handed helicity, 
\begin{equation}
\begin{split}
A_{\mu}^h(x)=\int \frac{{\rm d}^{3}p}{(2\pi)^{3}}\frac{1}{\sqrt{2 |{\bm p}|}}
\Bigl(a^{h}_{\bm{p}}\epsilon_{\mu}^{h}(p){\rm e}^{-\ri p\cdot x}+a^{h\dag}_{\bm{p}}\epsilon^{h*}_{\mu}(p){\rm e}^{\ri p\cdot x}
\Bigr),
\end{split}
\end{equation}
where $h = {\rm R, L}$ represents the index of helicity, 
and $a^{h}_{\bm{p}}$ and $a^{h\dag}_{\bm{p}}$ are annihilation and creation operators, respectively, that satisfy
the commutation relation 
\begin{equation}
\label{comm}
[a^h_{\bm p}, a_{{\bm p}'}^{h' \dag}] = (2\pi)^3 \delta^{h h'} \delta^{(3)}({\bm p} - {\bm p}').
\end{equation}

We are interested in the lesser and greater propagators for right/left-handed photons, 
which are defined as \cite{le2000thermal}
\begin{eqnarray}
G^{h<}_{\mu\nu}(q,X)=\int {\rm d}^4Y{\rm e}^{{\rm i}\frac{q\cdot Y}{\hbar}}\langle A^h_{\nu}(y)A^h_{\mu}(x)\rangle\,,
\qquad G^{h>}_{\mu\nu}(q,X)=\int {\rm d}^4Y{\rm e}^{{\rm i}\frac{q\cdot Y}{\hbar}}\langle A^h_{\mu}(x)A^h_{\nu}(y)\rangle\,,
\end{eqnarray}
respectively, where $Y=x-y$ and $X=(x+y)/2$. 
We may focus on the lesser propagator for right-handed photons,
\begin{eqnarray}\nonumber
G^{{\rm R}<}_{\mu\nu}(q,X)
&=&\int {\rm d}^4Y{\rm e}^{{\rm i}\frac{q\cdot Y}{\hbar}} 
\int \frac{{\rm d}^{3}p'}{(2\pi)^{3}}\frac{1}{\sqrt{2|{\bm p'}|}}\int \frac{{\rm d}^{3}p}{(2\pi)^{3}}\frac{1}{\sqrt{2|{\bm p}|}}
\\
&&\times\Big(
\langle a^{{\rm R}\dag}_{\bm{p}'}a^{\rm R}_{\bm{p}}\rangle\epsilon_{\mu}^{{\rm R}}(p)\epsilon^{{\rm R}*}_{\nu}(p'){\rm e}^{ -\ri p_{-}\cdot X-\ri p_{+}\cdot Y}
+\langle a^{\rm R}_{\bm{p}'}a^{{\rm R}\dagger}_{\bm{p}}\rangle\epsilon^{{\rm R}*}_{\mu}(p)\epsilon_{\nu}^{{\rm R}}(p'){\rm e}^{ +\ri p_{-}\cdot X+\ri p_{+}\cdot Y}
\Big)\,,
\end{eqnarray}
where $p_{+}=(p+p')/2$ and $p_{-}=p-p'$.
One can easily show that $G^{h<}_{\mu\nu}(q,X)$ is a Hermitian matrix 
according to its definition. 
Carrying out the Wigner transformation with the $p_\pm$ momenta, one finds
\begin{eqnarray}\nonumber
G^{{\rm R}<}_{\mu\nu}(q,X)
&&
=\pi\int\frac{{\rm d}^3p_-}{(2\pi)^3}  {\rm e}^{-{\rm i}p_-\cdot X}
\frac{1}{\left[\left(|{\bm q}|^2+\frac{|{\bm p_-}|^2}{4}\right)^2-({\bm q}\cdot{\bm p_-})^2\right]^{1/4}}
\\\nonumber
&&\quad \times
\Bigg[\epsilon_{\mu}^{\rm R}\left(q+\frac{p_-}{2}\right)\epsilon^{{\rm R}*}_{\nu}\left(q-\frac{p_-}{2}\right)\langle a^{{\rm R}\dagger}_{\bm q-\frac{\bm p_-}{2}}a^{{\rm R}}_{\bm q+\frac{\bm p_-}{2}}\rangle
\delta\left(q^0-p_+^{0}\right)
\\
&&\qquad +
\epsilon^{{\rm R}}_{\nu}\left(-q+\frac{p_-}{2}\right) \epsilon_{\mu}^{{\rm R}*}\left(-q-\frac{p_-}{2}\right)
\langle a^{{\rm R}}_{-\bm{q}+\frac{\bm p_-}{2}}a^{{\rm R}\dagger}_{-\bm{q}-\frac{\bm p_-}{2}}\rangle
\delta\left(q^0+p_+^{0}\right)
\Bigg]\,,
\label{eq:GR0}
\end{eqnarray}
where
\begin{eqnarray}
p_+^0=\frac{1}{2}\left(\ \Big|{\bm q}+\frac{\bm p_-}{2}\Big|+\Big|{\bm q}-\frac{\bm p_-}{2}\Big| \ \right)
\, , \qquad 
p_-^0= \Big|{\bm q}+\frac{\bm p_-}{2}\Big|-\Big|{\bm q}-\frac{\bm p_-}{2}\Big| .
\end{eqnarray} 
For brevity, we take $\hbar=1$ above.

In order to perform  the $p_-$ integral analytically,
we expand the integrand with respect to $ p_- $ and retain the terms up to $\mathcal{O}(p_-)$ such as
\begin{eqnarray}
\epsilon_{\mu}^{\rm R}\left(q+\frac{p_-}{2}\right)\epsilon^{{\rm R}*}_{\nu}\left(q-\frac{p_-}{2}\right) 
= \Pi_{\mu\nu}^{(0)}(q) + \frac{p_-^{\alpha}}{2} \Pi_{\mu\nu\alpha}^{(1)}(q) + \mathcal{O}(p_-^2)
\,,
\end{eqnarray}
where 
\begin{eqnarray}
\label{Pi}
\Pi_{\mu\nu}^{(0)}(q)\equiv\epsilon_{\mu}^{\rm R}(q)\epsilon^{{\rm R}*}_{\nu}(q)\,,
\qquad \Pi_{\mu\nu\alpha}^{(1)}(q)\equiv \big(\partial_{q^{\alpha}}\epsilon_{\mu}^{\rm R}(q)\big)\epsilon^{{\rm R}*}_{\nu}(q)-\epsilon_{\mu}^{\rm R}(q)\big(\partial_{q^{\alpha}}\epsilon^{{\rm R}*}_{\nu}(q)\big)\,.
\end{eqnarray}
We also have the expansions $p_+^0 \approx |{\bm q}|+\mathcal{O}(|{\bm p_-}|^2) $ and 
$p_-^0 \approx \bm q\cdot \bm{p}_{-} / |{\bm q}| +\mathcal{O}(|{\bm p_-}|^2) $.
In the end, this expansion provides us with the Wigner functions up to $\mathcal{O}(\hbar)$. 
Plugging those expressions into Eq.~(\ref{eq:GR0}), we then find  
\begin{eqnarray} \label{Gmunu_R}
&&G^{{\rm R}<}_{\mu\nu}(q,X)
\\\nonumber
&&\approx\pi\int\frac{{\rm d}^3p_- }{(2\pi)^3|{\bm q}|} {\rm e}^{-{\rm i}p_-\cdot X}
\bigg[\delta\left(q_0-|{\bm q}|\right)\left(\Pi_{\mu\nu}^{(0)}(q) +\frac{p_{-}^{\alpha}}{2}\Pi_{\mu\nu\alpha}^{(1)}(q) \right)
\langle a^{{\rm R}\dagger}_{\bm q-\frac{\bm p_-}{2}}a^{{\rm R}}_{\bm q+\frac{\bm p_-}{2}}\rangle
\\\nonumber
&& \hspace{4.2cm}
+\delta(q_0+|{\bm q}|) \left(\Pi_{\nu\mu}^{(0)}(-q) + \frac{p_{-}^{\alpha}}{2}\Pi_{\nu\mu\alpha}^{(1)}(-q) \right)
\langle a^{{\rm R}}_{-\bm q+\frac{\bm p_-}{2}}a^{{\rm R}\dagger}_{-\bm q-\frac{\bm p_-}{2}}\rangle
\bigg]
\\
&&=2\pi\delta(q^2)\left[\theta(q_0)\left(\Pi_{\mu\nu}^{(0)}(q)+\frac{\rm i}{2}\Pi_{\mu\nu\alpha}^{(1)}(q)\partial^{\alpha}\right)
-\theta(-q_0)\left(\Pi_{\nu\mu}^{(0)}(-q)+\frac{\rm i}{2}\Pi_{\nu\mu\alpha}^{(1)}(-q)\partial^{\alpha}\right)\right]\check{f}_{\rm R}(q,X)\,,
\nonumber 
\end{eqnarray}
where we dropped the $\mathcal{O}(|{\bm p}_-|^2)$ terms in the integrand except for those contributing to the distribution functions. 
In the first term, we introduced the distribution function for right-handed photons 
\begin{eqnarray}
\check{f}_{\rm R}(\bm q,X)\equiv \int\frac{{\rm d}^3p_-}{(2\pi)^3}\langle a^{{\rm R}\dagger}_{\bm q-\frac{\bm p_-}{2}}a^{{\rm R}}_{\bm q+\frac{\bm p_-}{2}}\rangle {\rm e}^{-{\rm i}p_-\cdot X}\,.
\end{eqnarray} 
In the second term, the commutation relation (\ref{comm}) leads to 
 \begin{eqnarray}
 \int\frac{{\rm d}^3p_-}{(2\pi)^3}{\rm e}^{-{\rm i}p_-\cdot X}\langle a^{{\rm R}}_{-\bm q+\frac{\bm p_-}{2}}a^{{\rm R}\dagger}_{-\bm q-\frac{\bm p_-}{2}}\rangle=1+\check{f}_{\rm R}(-\bm q,X)
 \, .
 \end{eqnarray}
Combining those two cases, the distribution function $ \check{f}_{\rm R}(q,X) $ 
is defined for the four-momentum $ q^\mu $ as 
 \begin{eqnarray}
 \check{f}_{\rm R}(q,X) =
 \begin{cases}
    \check{f}_{\rm R}(\bm q,X) & (q_0=|\bm q|) \\
    -\left[1+\check{f}_{\rm R}(-\bm q,X) \right] & (q_0=-|\bm q|)\,.
  \end{cases}
  \end{eqnarray}
Note that the $\theta(-q_0)$ part in Eq.~(\ref{Gmunu_R}) characterizes the out-going photons.

To derive explicit forms of $\Pi_{\mu\nu}^{(0)}$ and $\Pi_{\mu\nu\alpha}^{(1)}$, 
it is crucial to use the expressions of $\epsilon^{h}_{\mu}$ from the spinor-helicity formalism in Sec.~\ref{sec_spinor_helicity}. 
Inserting Eq.~(\ref{epsilon^R}) into Eq.~(\ref{Pi}), we have
\begin{eqnarray}
&&\Pi_{\mu\nu}^{(0)}(q)=\frac{1}{2}\text{tr} \left(c_{\rm R}^{(-)} c_{\rm R}^{(-)\dagger} \sigma_{\mu}c_{\rm R}^{(+)} c_{\rm R}^{(+)\dagger} \sigma_{\nu}\right)\,, 
\\
&&  \Pi_{\mu\nu\alpha}^{(1)}(q)={\rm i}\ \text{tr}\Big[{\cal A}\Big(c_{\rm R}^{(-)} \partial_{q^{\alpha}}c_{\rm R}^{(-)\dagger} \Big)\sigma_{\mu}c_{\rm R}^{(+)} c_{\rm R}^{(+)\dagger} \sigma_{\nu}
-c_{\rm R}^{(-)} c_{\rm R}^{(-)\dagger} \sigma_{\mu}{\cal A}\Big(c_{\rm R}^{(+)}\partial_{q^{\alpha}}c_{\rm R}^{(+)\dagger}\Big)\sigma_{\nu}
\Big]\,.
\end{eqnarray}
Here and below, we define ${\cal A} (M) = - {\rm i} (M - M^\dagger)/2 $ for a matrix-valued quantity $M$ and 
omit the arguments as $c_{\rm R}^{(\pm)} = c_{\rm R}^{(\pm)} (q)$ and $c_{\rm R}^{(\pm)\dagger} = c_{\rm R}^{(\pm)\dagger}(q)$.
Now the computations of $\Pi_{\mu\nu}^{(0)}$ and $\Pi_{\mu\nu\alpha}^{(1)}$ are straightforward, 
and the details are given in Appendix~\ref{app_Pi}.
The results are found to be 
\begin{eqnarray}
\label{Pimunu0}
&&\Pi_{\mu\nu}^{(0)}(q)=\frac{1}{2}\big[(n_{\mu}n_{\nu}-\eta_{\mu\nu})-\hat{q}_{\perp\mu}\hat{q}_{\perp\nu}
-{\rm i}\epsilon_{\mu\nu\alpha\beta}n^{\alpha}\hat{q}^{\beta}\big]\,,
\\
\label{Pimunu1}
&&\Pi_{\mu\nu\alpha}^{(1)}(q)=2{\rm i}\Big[a^-_{\alpha}(q)-a^+_{\alpha}(q)\Big]\Pi^{(0)}_{\mu\nu}(q)
-\frac{1}{2|{\bm q}|}\Big[{\rm i}\hat{q}_{\perp(\mu}\epsilon_{\nu)\alpha\beta\rho}n^{\beta}\hat{q}^{\rho}
+\hat{q}_{\perp[\mu}\big(\eta_{\nu]\alpha}-n_{\nu]}n_{\alpha}\big)\Big]\,,
\end{eqnarray}
where we have  the Berry connections 
\begin{eqnarray}
\label{Berry}
a^{\mu}_{\pm}(q) \equiv {\rm i}c_{\rm R}^{(\pm)\dagger} \partial_{q_{\mu}} c_{\rm R}^{(\pm)}
\, .
\end{eqnarray}
Note that $a_{\pm}^{\mu}(-q)=-a_{\mp}^\mu(q)$. 
Since $\Pi_{\nu\mu}^{(0)}(-q)=\Pi_{\mu\nu}^{(0)}(q)$ and $\Pi_{\nu\mu\alpha}^{(1)}(-q)=\Pi_{\mu\nu\alpha}^{(1)}(q)$, 
the Hermitian property of Wigner functions is maintained properly. 
According to these results, Eq.~(\ref{Gmunu_R}) reads 
\begin{eqnarray}\nonumber\label{Gmunu_R_exp}
&&G^{{\rm R}<}_{\mu\nu}(q,X)
\\\nonumber
&&=2\pi\delta(q^2)\text{sgn}(q_0)
\Bigg[ \Big( \text{Re} [\Pi_{\mu\nu}^{(0)}(q)]-\frac{\hbar}{2}\text{Im} [\Pi_{\mu\nu\alpha}^{(1)}(q)] \partial^{\alpha} \Big)
+{\rm i}\Big( \text{Im}[\Pi_{\mu\nu}^{(0)}(q)]+\frac{\hbar}{2}\text{Re}[\Pi_{\mu\nu\alpha}^{(1)}(q)]\partial^{\alpha}\Big)\Bigg]\check{f}_{\rm R}(q,X)
\\
&&=\pi\delta(q^2)\text{sgn}(q\cdot n)\Bigg[\bigg(P^{(n)}_{\mu\nu}-\frac{\hbar q_{\perp(\mu}S^{(n)}_{\nu)\alpha}\partial^{\alpha}}{2(q\cdot n)^2}
\bigg)
-{\rm i}\bigg( S_{\mu\nu}^{(n)}+\frac{\hbar q_{\perp[\mu}\partial_{\perp\nu]}}{2(q\cdot n)^2}\bigg)
\Bigg] {f}_{\rm R}(q,X)
\end{eqnarray}
up to $\mathcal{O}(\hbar)$. Here, the standard polarization tensor in the Coulomb gauge 
and the spin tensor of photons are, respectively, given as 
\begin{equation}
P^{(n)}_{\mu\nu} = n_{\mu}n_{\nu}-\eta_{\mu\nu}-\hat{q}_{\perp\mu}\hat{q}_{\perp\nu}\,, \qquad 
S_{\mu\nu}^{(n)}=\frac{\epsilon_{\mu\nu\alpha\beta}q^{\alpha}n_{\beta}}{q\cdot n}
\, .
\end{equation}
Absorbing the Berry connections, we have defined a frame-dependent distribution function 
\begin{eqnarray}
{f}_{\rm R}(q,X)\equiv \Big(1+\hbar\big[a_{+}^{\alpha}(q)-a_-^{\alpha}(q)\big] \partial_{\alpha}\Big) \check{f}_{\rm R}(q,X)
\, .
\end{eqnarray}
In fermionic systems, it has been shown that 
such a quantum correction is responsible for the so-called side jump effect \cite{Chen:2015gta,Hidaka:2016yjf}.

Based on Eq.~(\ref{Gmunu_R_exp}), we may also write down the lesser propagator for left-handed photons 
\begin{eqnarray}
G^{{\rm L}<}_{\mu\nu}(q,X)
=\pi\delta(q^2)\text{sgn}(q\cdot n)\Bigg[\bigg(P^{(n)}_{\mu\nu} + \frac{\hbar q_{\perp(\mu}S^{(n)}_{\nu)\alpha}\partial^{\alpha}}{2(q\cdot n)^2}
\bigg)
+{\rm i}\bigg(S_{\mu\nu}^{(n)}-\frac{\hbar q_{\perp[\mu}\partial_{\perp\nu]}}{2(q\cdot n)^2}\bigg)
\Bigg]{f}_{\rm L}(q,X)
\label{Gmunu_L_exp}
\,.
\end{eqnarray} 
Combining $G^{{\rm R}<}_{\mu\nu}$ and $G^{{\rm L}<}_{\mu\nu}$, we obtain the full lesser propagator for photons 
\begin{eqnarray}\nonumber\label{Gmunu_full}
G^{<}_{\mu\nu}&\equiv& G^{{\rm R}<}_{\mu\nu}+G^{{\rm L}<}_{\mu\nu}
\\
&=&2\pi\delta(q^2)\text{sgn}(q\cdot n)\Bigg[\bigg(P^{(n)}_{\mu\nu} {f}_{\rm V}
-\frac{\hbar q_{\perp(\mu}S^{(n)}_{\nu)\alpha}\partial^{\alpha}}{2(q\cdot n)^2}{f}_{\rm A}
\bigg)
-{\rm i}\bigg(S_{\mu\nu}^{(n)}{f}_{\rm A}+
\frac{\hbar q_{\perp[\mu}\partial_{\perp\nu]}}{2(q\cdot n)^2}{f}_{\rm V}\bigg)
\Bigg]\,,
\end{eqnarray}
where ${f}_{\rm A}\equiv ({f}_{\rm R}-{f}_{\rm L})/2$ and ${f}_{\rm V}\equiv ({f}_{\rm R}+{f}_{\rm L})/2$ in our conventions. 
Note that the latter definition of ${f}_{\rm V}$ is determined by the standard lesser propagator of photons at $\mathcal{O}(\hbar^0)$ 
so that ${f}_{\rm V}$ reduces to the distribution function of unpolarized photons. 
Although $G^{<}_{\mu\nu}$ itself is a gauge-dependent quantity, 
it can be utilized to calculate gauge-invariant quantities such as the energy-momentum tensor shown in Appendix.~\ref{app_EM_tensor}.
For the greater propagator, we simply have to replace ${f}_{\rm V}$ by $(1+{f}_{\rm V})$ with keeping ${f}_{\rm A}$ unchanged. More precisely, we have
\begin{equation}
G^{>}_{\mu\nu}=2\pi\delta(q^2)\text{sgn}(q\cdot n)\Bigg[\bigg(P^{(n)}_{\mu\nu}(1+{f}_{\rm V})
-\frac{\hbar q_{\perp(\mu}S^{(n)}_{\nu)\alpha}\partial^{\alpha}}{2(q\cdot n)^2}{f}_{\rm A}
\bigg)
-{\rm i}\bigg(S_{\mu\nu}^{(n)}{f}_{\rm A}+
\frac{\hbar q_{\perp[\mu}\partial_{\perp\nu]}}{2(q\cdot n)^2}{f}_{\rm V}\bigg)
\Bigg]\,.
\end{equation}
Since we will always work in the frame $n^{\mu}= (1, {\bm 0})$, we hereafter omit the superscript ``${(n)}$" for the polarization tensor $P^{(n)}_{\mu\nu}$ and the spin tensor $S^{(n)}_{\mu\nu}$ in the following.
We will also attach an index ``$\gamma$" to the distribution function like $f^\gamma$ 
and similarly $S_{\mu \nu}^\gamma$ for the spin tensor to stress that these are of photons.

Notably, as opposed to the approach of solving the Wigner function from the equations of motion in Ref.~\cite{Huang:2020kik}, here the Wigner function is explicitly derived as an expectation value of the corresponding operator in the quantum field theory. Although two results agree, the free function $U$ appearing in Ref.~\cite{Huang:2020kik} is uniquely fixed in the present calculation, which further clarifies the physical picture underlying these results.

\section{Helicity currents and chiral/Zilch vortical effect}
\label{sec_CS_current}
The primary purpose of this section is to reproduce the CVE/ZVE for photons found in the previous literature such as Ref.~\cite{Huang:2020kik} to ensure the validity of our formalism.
We may now utilize $G^{h<}_{\mu\nu}$ to evaluate the CS currents for polarized photons,
\begin{eqnarray}
\label{def_CS_current}
\mathcal{K}^{\mu}_{h}(x)\equiv A^h_{\nu}(x)\tilde{F}^{\mu\nu}_h(x)=\frac{1}{2}\epsilon^{\mu\nu\alpha\beta}A^h_{\nu}(x)\overleftrightarrow{\partial}_{\alpha}A^h_{\beta}(x)\,, 
\end{eqnarray}
where $\tilde{F}^{\mu\nu}_h = \epsilon^{\mu \nu \alpha \beta} F^h_{\alpha \beta}/2$ and $\overleftrightarrow{\partial}_{\alpha}\equiv\overrightarrow{\partial}_{\alpha}-\overleftarrow{\partial}_{\alpha}$. 
The derivative $\overleftrightarrow{\partial}_{\alpha}$ acting on both $A_{\nu}^h(x)$ and $A_{\beta}^h(x)$ is convenient for computations in terms of Wigner functions as in the case for energy-momentum tensors of fermions discussed in, e.g., Ref.~\cite{Yang:2018lew}. In light of Eq.~(\ref{def_CS_current}), the CS currents can be derived from the Wigner functions as
\begin{eqnarray}
\mathcal{K}^{\mu}_{h}(X)&=&\frac{1}{2}\epsilon^{\mu\nu\alpha\beta}
\int\frac{{\rm d}^4q}{(2\pi)^4}\int {\rm d}^4Y
{\rm e}^{{\rm i}\frac{q\cdot Y}{\hbar}}\left[\langle A^h_{\nu}(y)\partial_{x^\alpha}A^h_{\beta}(x)\rangle-\langle \big(\partial_{y^\alpha}A^h_{\nu}(y)\big)A^h_{\beta}(x)\rangle \right]
\nn\\
&=&-\frac{\rm i}{\hbar}\int\frac{{\rm d}^4q}{(2\pi)^4} \epsilon^{\mu\nu\alpha\beta}q_{\alpha}G^{h<}_{\beta\nu}(q,X)\,.
\label{CS_Wigner}
\end{eqnarray}
 Apparently, only the anti-symmetric and imaginary component of Wigner functions contributes to the CS currents. Substituting Eqs.~(\ref{Gmunu_R_exp}) and (\ref{Gmunu_L_exp}) into Eq.~(\ref{CS_Wigner}), the CS currents read
\begin{eqnarray}\nonumber
\mathcal{K}^{\mu}_{h}(X)&=&\mp \frac{\pi}{\hbar}
\int\frac{{\rm d}^4q}{(2\pi)^4}
\delta(q^2)\text{sgn}(q\cdot n)\epsilon^{\mu\nu\alpha\beta}q_{\alpha}\left(S^{\gamma}_{\beta\nu}\pm\frac{\hbar}{(q\cdot n)^2}q_{\perp\beta}\partial_{\perp\nu}\right){f}_{h}^{\gamma}(q,X)
\\
&=&\int\frac{{\rm d}^4q}{(2\pi)^3\hbar}\delta(q^2)\text{sgn}(q\cdot n)\Big(\pm q^{\mu}+ \frac{\hbar}{2} S_{\gamma}^{\mu\nu} \partial_{\nu}\Big){f}_{h}^{\gamma}(q,X)\,.
\end{eqnarray}
Accordingly, the full CS current becomes
\begin{eqnarray}
\mathcal{K}^{\mu}(X)\equiv\mathcal{K}^{\mu}_{{\rm R}}(X)+\mathcal{K}^{\mu}_{{\rm L}}(X)
=\int\frac{{\rm d}^4q}{(2\pi)^3\hbar}\delta(q^2)\text{sgn}(q\cdot n)\hat{\mathcal{K}}^{\mu}(q,X)\,,
\end{eqnarray}
where the phase-space CS current density is given as 
\begin{eqnarray}\label{CS_current_density}
\hat{\mathcal{K}}^{\mu}(q,X)\equiv \frac{-{\rm i}\epsilon^{\mu\nu\alpha\beta}q_{\nu}\text{Im}\big(G^{<}_{\alpha\beta}(q,X)\big)}{2\pi\sgn(q\cdot n)\delta(q^2)}=
2 q^{\mu}{f}_{\rm A}^{\gamma}+\hbar S_{\gamma}^{\mu\nu}\partial_{\nu}{f}_{\rm V}^{\gamma}
\, .
\end{eqnarray}
Despite the gauge dependence, the CS current could be regarded as the spin component of the angular momentum for photons \cite{Leader:2013jra, Fukushima:2020qta}. 
Spin-one nature of photons manifests itself in the side-jump term associated with $S_{\gamma}^{\mu\nu}$ in Eq.~(\ref{CS_current_density}) that is twice larger than the corresponding term in
the axial charge current characterizing the spin polarization of massless fermions.

Nevertheless, due to the gauge dependence of the CS current, it is tempting to also introduce the zilch current as a gauge-invariant quantity delineating the helicity currents of photons \cite{Lipkin_Zilch,morgan1964two,Kibble_Zilch,PhysRevLett.104.163901}. We may make comparisons with the related studies of the zilch vortical effect in Refs.~\cite{Chernodub:2018era,Copetti:2018mxw,Huang:2020kik}. 
Let us now consider the spin-3 zilch
\begin{eqnarray}
Z_{\mu\nu\rho}\equiv
\frac{1}{2}\Big[F_{\mu}^{\,\,\alpha}\partial_{\rho}\tilde{F}_{\nu\alpha}
-(\partial_{\rho}F_{\nu}^{\,\,\alpha})\tilde{F}_{\mu\alpha}\Big]\,.
\end{eqnarray}
To construct the zilch in terms of the Wigner function, we rewrite the zilch current with the gauge field 
\begin{eqnarray}
\nonumber
Z_{\mu\nu\rho}(x,y)
&=&
\frac{1}{4}(\partial_{x\rho}-\partial_{y\rho})\epsilon_{(\mu\alpha\lambda\sigma}
\big[\partial_{y\nu)}\partial_{x}^{\lambda} A^{\alpha}(y)A^{\sigma}(x)
+\partial_{y}^{\alpha}\partial_{x}^{\sigma}A_{\nu)}(y)A^{\lambda}(x)
\big]
\\
&&-
\frac{1}{4}(\partial_{x\rho}+\partial_{y\rho})\epsilon_{[\mu\alpha\lambda\sigma}
\big[\partial_{y\nu]}\partial_{x}^{\lambda} A^{\alpha}(y)A^{\sigma}(x)
+\partial_{y}^{\alpha}\partial_{x}^{\sigma}A_{\nu]}(y)A^{\lambda}(x)
\big]
\,,
\end{eqnarray}
which yields
\begin{eqnarray}\nonumber
&&\langle Z_{\mu\nu\rho}\rangle (q,X)
\\\nonumber
&&=-\frac{{\rm i}q_{\rho}}{2 \hbar}
\epsilon_{(\mu\alpha\lambda\sigma}
\Big[{\rm D}_{\nu)}{\rm D}^{*\lambda} G^{<\sigma\alpha}(q,X)
+{\rm D}^{\alpha}{\rm D}^{*\sigma} G^{<\lambda}_{\quad\nu)}(q,X)
\Big]
\\\nonumber
&&\quad- \frac14\epsilon_{[\mu\alpha\lambda\sigma}\partial_{\rho}
\Big[{\rm D}_{\nu]}{\rm D}^{*\lambda}G^{<\sigma\alpha}(q,X)
+{\rm D}^{\alpha}{\rm D}^{*\sigma} G^{<\lambda}_{\quad \nu]}(q,X)
\Big]
\\\nonumber
&&=-\frac{{\rm i}q_{\rho}}{2\hbar^3}
\epsilon_{(\mu\alpha\lambda\sigma}
\Big[\Big(q_{\nu)}q^{\lambda}+\frac{{\rm i}\hbar}{2}\big(q_{\nu)}\partial^{\lambda}-q^{\lambda}\partial_{\nu)}\big)\Big) G^{<\sigma\alpha}
+\frac{{\rm i}\hbar}{2}\big(q^{\alpha}\partial^{\sigma}-q^{\sigma}\partial^{\alpha}\big)  G^{<\lambda}_{\quad\nu)}
+\mathcal{O}(\hbar^2)
\Big]
\\
&&\quad- \frac{\epsilon_{[\mu\alpha\lambda\sigma}}{4\hbar^2}\partial_{\rho}
\Big[\Big(q_{\nu]}q^{\lambda}+\frac{{\rm i}\hbar}{2}\big(q_{\nu]}\partial^{\lambda}-q^{\lambda}\partial_{\nu]}\big)\Big) G^{<\sigma\alpha}
+\frac{{\rm i}\hbar}{2}\big(q^{\alpha}\partial^{\sigma}-q^{\sigma}\partial^{\alpha}\big)  G^{<\lambda}_{\quad\nu]}
+\mathcal{O}(\hbar^2)
\Big]
\,,
\end{eqnarray}
where ${\rm D}_{\nu}\equiv {\partial_{\nu}}/{2}+{\rm i}{q_{\nu}}/{\hbar}$. 

Although the general form of the zilch is involved, it may be simplified with the aid of a specific power counting. 
Analogous to the power-counting scheme proposed in Ref.~\cite{Yang:2020hri} for fermions, 
we assume that ${f}_{\rm V}^{\gamma} = \mathcal{O}(\hbar^0)$ and ${f}_{\rm A}^{\gamma} = \mathcal{O}(\hbar^1)$. 
We confirm, {\it a posteriori}, that these assumptions are satisfied 
when a nonzero ${f}_{\rm A}^{\gamma}$ is induced by a vorticity in local thermal equilibrium (see below).
When this $  \hbar$ counting is applied, the leading-order contribution to the zilch current reads
\begin{eqnarray}\label{Zilch_spin3}\nonumber
\langle Z_{\mu\nu\rho}\rangle (q,X)&\approx& 
{\pi} \frac{q_{\rho}}{\hbar^3}\sgn(q_0)\delta(q^2)\left(\hat{\mathcal{K}}_{q(\mu}q_{\nu)}
+\hbar\epsilon_{(\mu\alpha\lambda\sigma}P^{\lambda}_{\nu)}q^{\alpha}\partial^{\sigma}{f}_{\rm V}^{\gamma}\right)
\\
&=& {2\pi} \frac{q_{\rho}}{\hbar^3} \sgn(q_0)\delta(q^2)\left(2q_{\mu}q_{\nu}{f}_{\rm A}^{\gamma}+\hbar q_{(\mu}S^{\gamma}_{\nu)\sigma}\partial^{\sigma}{f}_{\rm V}^{\gamma}
\right)\,,
\end{eqnarray}
where $q_0=q\cdot n$ and we use subindices such as $\mathcal{F}_{q}$ to represent an arbitrary function $\mathcal{F}(q,X)$ in the phase space.  
The zilch current can be defined as $Z^{\alpha}(X)\equiv \int {\rm d}^4q\hat{Z}^{\alpha}(q,X)/(2\pi)^4$ with
\begin{eqnarray}
\label{eq:zilch-density}
\hat{Z}^{\alpha}(q,X)\equiv \Theta^{\alpha\mu}_{(n)}n^{\nu}n^{\rho}\langle Z_{\mu\nu\rho}\rangle (q,X)
=2\pi \frac{q_0^2}{\hbar^3} \sgn(q_0)\delta(q^2)\big(2q^{\alpha}_{\perp}{f}_{\rm A}^{\gamma}+\hbar S_{\gamma}^{\alpha\sigma}\partial_{\sigma}{f}_{\rm V}^{\gamma}\big)\,,
\end{eqnarray}
where $\Theta_{(n)}^{\alpha\mu}\equiv \eta^{\alpha\mu}-n^{\alpha}n^{\mu}$. 

We may further investigate the CS and zilch currents in a local thermal equilibrium with nonzero fluid vorticity $\omega^{\mu}=\epsilon^{\mu\nu\alpha\beta}u_\nu \partial_\alpha u_\beta/2$, where $u^{\mu}$ is the fluid four velocity. 
Although it is in principle not possible to determine the equilibrium distribution functions without solving the kinetic theory with collisional effects, we may physically expect that the equilibrium distribution functions take the Bose-Einstein form with the $\hbar$ correction due to the spin-vorticity coupling 
\begin{eqnarray}
{f}_{h, \text{eq}}^{\gamma}=\frac{1}{{\rm e}^{g_{h}}-1}\,,\qquad g_{\rm R/L}=q \cdot U \pm \frac{\hbar \lambda_{\gamma}}{2}S_{\gamma}^{\mu\nu}\Omega_{\mu\nu}\,, 
\end{eqnarray} 
where $U^{\mu}=\beta u^{\mu}$ with $\beta = 1/T$ and $T$ being a temperature. 
The spin-vorticity coupling term is given by the helicity of photons $\lambda_{\gamma}$ 
and the thermal vorticity $\Omega_{\mu\nu}\equiv\partial_{[\mu}U_{\nu]}/2$. 
This yields the leading-order terms in $ \hbar $: 
\begin{eqnarray}\label{assumption_equil_fA_text}
{f}^{\gamma}_{{\rm A}, \text{eq}}=-\hbar \lambda_{\gamma}\hat{q}_{\mu}\tilde{\Omega}^{\mu\nu}n_{\nu}N(q\cdot u)\big[1+N(q\cdot u)\big]\,,
\qquad 
{f}^{\gamma}_{{\rm V},\text{eq}} = N(q\cdot u)\,,
\end{eqnarray}     
where $N(q\cdot u) \equiv \left({\rm e}^{\beta q\cdot u}-1\right)^{-1}$
and $\tilde{\Omega}_{\rho\sigma}\equiv \epsilon_{\rho\sigma\alpha\beta}\Omega^{\alpha\beta}/2$. 
Indeed, these leading-order expressions with the spin-vorticity coupling meet our assumptions that 
${f}_{\rm V}^{\gamma} = \mathcal{O}(\hbar^0)$ and ${f}_{\rm A}^{\gamma} = \mathcal{O}(\hbar^1)$, 
unless the magnitude of the thermal vorticity compensates the smallness of $  \hbar$. 
In more general cases, a stronger thermal vorticity could induce a larger $ {f}_{\rm A}^{\gamma} $ beyond our counting scheme. 

By analogy to the case of fermions where the helicity is $\lambda_{\rm f} = 1/2$ \cite{Chen:2015gta,Hidaka:2016yjf}, 
we may anticipate $\lambda_{\gamma}=1$ for photons. 
As shown around Eq.~(\ref{eq:photon-helicity}), we derive $\lambda_{\gamma}=1$ by demanding a frame independence of the zilch. 
Using Eq.~(\ref{assumption_equil_fA_text}), the zilch current in the equilibrium state is given by
\begin{eqnarray}\nonumber
Z^{\alpha}_{\text{eq}}(X)
&=&\int \frac{{\rm d}^4q}{(2\pi)^4}\hat{Z}^{\alpha}_{\text{eq}}(q,X)
\\\nonumber
&=&-\int \frac{{\rm d}^4q}{(2\pi)^3\hbar^2} q_0\sgn(q_0)\delta(q^2)\left(2q^{\alpha}_{\perp}\lambda_{\gamma}q_{\rho}\tilde{\Omega}^{\rho\sigma}n_{\sigma}+q_0S_{\gamma}^{\alpha\sigma}\Omega_{\sigma\rho}q^{\rho}\right)N(q \cdot u )\big[1+N(q \cdot u )\big]
\\\nonumber
&=&(2\lambda_{\gamma}+2)\frac{\tilde{\Omega}^{\alpha\sigma}n_{\sigma}}{6\pi^2\hbar^2}\int_{0}^{\infty} {\rm d}|\bm q||\bm q|^4N(|\bm q|)\big[1+N(|\bm q|)\big]
\\
&=&(\lambda_{\gamma}+1)\frac{4\pi^2}{45\hbar^2} T^4\omega^{\alpha}\,.
\end{eqnarray}
Here, we assume a small flow velocity and thus $u^{\mu} \approx n^{\mu}$. 
Then, we used $\tilde{\Omega}^{\alpha\sigma}n_{\sigma} \approx \beta \omega^{\alpha}$
which follows from
\begin{eqnarray}
\tilde{\Omega}^{\mu\nu} = \frac12 \epsilon^{\mu\nu\lambda\rho} \left ( \beta \partial_{\lambda} u_{\rho} - u_{\lambda} \partial_{\rho} \beta \right)
\, .
\end{eqnarray}
When taking $\lambda_{\gamma}=1$ as determined just below, one finds 
\begin{eqnarray}
Z^{\alpha}_{\text{eq}}(X)=\frac{8\pi^2}{45\hbar^2} T^4\omega^{\alpha}\,.
\end{eqnarray}
This equilibrium zilch current agrees with the results in Refs. \cite{Chernodub:2018era,Copetti:2018mxw,Huang:2020kik}. 
On the other hand, for the equilibrium CS current, one finds
\begin{eqnarray}\nonumber
\mathcal{K}^{\mu}_{\text{eq}}&=&-\int\frac{{\rm d}^4q}{(2\pi)^3}\text{sgn}(q\cdot n)\delta(q^2)\left(2q^{\mu}\lambda_{\gamma}\frac{q^{\rho}\tilde{\Omega}_{\rho\sigma}n^{\sigma}}{q\cdot n}+S_{\gamma}^{\mu\nu}\Omega_{\nu\sigma}q^{\sigma}\right)N(q\cdot n)\big[1+N(q\cdot n)\big]
\\
&=&\frac{\lambda_{\gamma}+1}{9}T^2 \omega^{\mu}.
\end{eqnarray}
When taking $\lambda_{\gamma}=1$, the CS current is also consistent with the one in Ref.~\cite{Huang:2020kik}. 

In the following, we show how to obtain $\lambda_{\gamma}=1$ by demanding 
a frame independence of the zilch current (\ref{Zilch_spin3}). 
Note that it is more plausible to use the gauge-independent zilch current instead of the gauge-dependent CS current.
Inserting Eq.~(\ref{assumption_equil_fA_text}) into Eq.~(\ref{Zilch_spin3}), we have 
\begin{eqnarray}\nonumber\label{Zilch_equ_check}
2q_{\mu}q_{\nu}{f}_{\rm A}^{\gamma}+\hbar q_{(\mu}S^{\gamma}_{\nu)\sigma}\partial^{\sigma}{f}_{\rm V}^{\gamma}
&=& -\hbar \left(2q_{\mu}q_{\nu}\frac{\lambda_{\gamma} q_{\rho}\tilde{\Omega}^{\rho\sigma}n_{\sigma}}{q\cdot n}
+\frac{q_{(\mu}\epsilon_{\nu)\sigma\alpha\beta}q^{\alpha}n_{\beta}}{q\cdot n}\Omega^{\sigma\kappa}q_{\kappa}\right)N(q \cdot u )\big[1+N(q \cdot u )\big]
\\
&=&-\hbar \left(2(\lambda_{\gamma}-1) q_{\mu}q_{\nu}\frac{q_{\rho}\tilde{\Omega}^{\rho\sigma}n_{\sigma}}{q\cdot n}
-q_{(\mu}\tilde{\Omega}_{\nu)\alpha}q^{\alpha}\right)N(q \cdot u )\big[1+N(q \cdot u )\big],
\end{eqnarray}
where we used the Schouten identity
\begin{eqnarray}\nonumber
q_{\mu}\epsilon_{\nu\sigma\alpha\beta}q_{\kappa}
\frac{q^{\alpha}n_{\beta}\Omega^{\sigma\kappa}}{2q\cdot n}&=&
q_{\mu}\big(q_{\nu}\epsilon_{\kappa\sigma\alpha\beta}+q_{\sigma}\epsilon_{\nu\kappa\alpha\beta}+q_{\alpha}\epsilon_{\nu\sigma\kappa\beta}+q_{\beta}\epsilon_{\nu\sigma\alpha\kappa}\big)\frac{q^{\alpha}n_{\beta}\Omega^{\sigma\kappa}}{2q\cdot n}
\\
&=&-q_{\mu}q_{\nu}\frac{q_{\rho}\tilde{\Omega}^{\rho\sigma}n_{\sigma}}{q\cdot n}
-\frac{q_{\mu}\epsilon_{\nu\sigma\alpha\beta}q^{\alpha}n_{\beta}\Omega^{\sigma\kappa}q_{\kappa}}{2q\cdot n}
-q_{\mu}\tilde{\Omega}_{\nu\alpha}q^{\alpha}\,.
\end{eqnarray}
For the frame-dependent term in Eq.~(\ref{Zilch_equ_check}) to be absent, 
the helicity parameter is required to be 
\begin{eqnarray}
\label{eq:photon-helicity}
 \lambda_{\gamma}=1
 \, .
\end{eqnarray} 
Then, the equilibrium zilch current should read 
\begin{eqnarray}
\label{eq:Z_eq}
\langle Z_{\mu\nu\rho}\rangle_{\text{eq}} (q,X)
&=&2\pi \frac{q_{\rho}}{\hbar^3}\sgn(q_0)\delta(q^2)q_{(\mu}\tilde{\Omega}_{\nu)\alpha}q^{\alpha}N(q \cdot u )\big[1+N(q \cdot u )\big]\, .
\end{eqnarray}
In fact, the frame and gauge invariances of the zilch current $\langle Z_{\mu\nu\rho}\rangle_{\text{eq}} (q,X)$ can be succeeded to $\hat{Z}^{\alpha}_{\text{eq}}(q,X)\equiv \Theta^{\alpha\mu}_{(u)}{u}^{\nu}{u}^{\rho}\langle Z_{\mu\nu\rho}\rangle_{\text{eq}} (q,X)$ if one defines the projection with the flow vector $u^{\mu}$ instead of that in Eq.~(\ref{eq:zilch-density}) with $ n^\mu $. Nevertheless, the final form of the equilibrium zilch current remains the same 
within the current working regime where $u^\mu \approx n^\mu $. In summary, the CVE and ZVE for photons are successfully re-derived from our approach.

\section{Quantum Kinetic theory for Photons}\label{sec_QKT_photons}
In this section, we would like to construct the QKT for on-shell photons. We may first start with the free-streaming case for simplicity as also shown in Ref.~\cite{Huang:2020kik} and then move to a more sophisticated construction incorporating collisions. The primary goal is to derive the generic form of collision term with quantum correction at $\mathcal{O}(\hbar^1)$ in terms of lesser/greater self-energies. In such a formalism as recently constructed for fermions \cite{Hidaka:2016yjf,Yang:2020hri}, given the information of classical collision term at $\mathcal{O}(\hbar^0)$ from, e.g., diagrammatic calculations, one can systematically obtain the corresponding quantum correction in collisions.%
\footnote{A concrete example for the application of QKT has been shown in Ref.~\cite{Yamamoto:2020zrs} on the neutrino transport in core-collapse supernovae. Given the neutrino self-energies at $\mathcal{O}(\hbar^0)$ obtained through the weak interaction with thermal nucleons, the $\mathcal{O}(\hbar)$ correction in the collision term of neutrino QKT is derived, which explicitly reveals the influence from vorticity and magnetic fields.}      
\subsection{Free-streaming case}
Considering the equation of motion for free photons, one finds 
\begin{eqnarray}
\Big(\eta^{\mu}_{\mbox{ }\rho}\partial^2-\partial_{\rho}\partial^{\mu}\Big)_{x}\langle A^{\nu}_h(y)A^{\rho}_h(x) \rangle=0\,,
\end{eqnarray}
which results in
\begin{eqnarray}\label{EOM_WT}
&&\Bigg[\frac{1}{4}(\eta^{\mu}_{\mbox{ }\rho}\partial^2-\partial_{\rho}\partial^{\mu})-\frac{1}{\hbar^2}(\eta^{\mu}_{\mbox{ }\rho}q^2-q_{\rho}q^{\mu})-\frac{\rm i}{2\hbar}(2\eta^{\mu}_{\mbox{ }\rho}q\cdot\partial-q_{\rho}\partial^{\mu}-q^{\mu}\partial_{\rho})\Bigg]G^{h<\rho\nu}(q,X)=0\,,
\end{eqnarray}
after the Wigner transformation. 
We may decompose the lesser propagator into the symmetric and anti-symmetric parts, 
\begin{eqnarray}
G^{h<\rho\nu}=G^{h<\rho\nu}_{\rm S}+{\rm i}G^{h<\rho\nu}_{\rm A}\,, \quad 
G^{h<\rho\nu}_{\rm S}\equiv\frac{1}{2}G^{h<(\rho\nu)}\,, \quad 
G^{h<\rho\nu}_{\rm A}\equiv -\frac{{\rm i}}{2}G^{h<[\rho\nu]}\,.
\end{eqnarray} 
According to the Hermitian property of photon Wigner functions, 
$G^{h<(\rho\nu)}$ and $G^{h<[\rho\nu]}$ are purely real and imaginary, respectively. 
In the Coulomb gauge such that $\partial_{x\perp}^{\mu}\langle A_{\nu}(y)A_{\mu}(x) \rangle=0$ and $\partial_{y\perp}^{\nu}\langle A_{\nu}(y)A_{\mu}(x) \rangle=0$, we should have the following constraints for Wigner functions in phase space, 
\begin{eqnarray}
\label{constraint}
\left(q_{\perp\rho}+\frac{{\rm i}\hbar}{2}\partial_{\perp\rho}\right)G^{<\rho\nu}=0\,, \qquad 
n_{\rho}G^{<\rho\nu}=0\,,
\end{eqnarray}
and hence,
\begin{eqnarray}\label{constraint_Coulomb}
2q_{\perp\rho}G^{h\rho\nu}_{\rm S}-\hbar\partial_{\perp\rho}G^{h\rho\nu}_{\rm A}=0,\qquad 
\hbar\partial_{\perp\rho}G^{h\rho\nu}_{\rm S}+2q_{\perp\rho}G^{h\rho\nu}_{\rm A}=0, 
\quad n_{\rho}G^{h<\rho\nu}_{\rm S/A}=G^{h <\rho\nu}_{\rm S/A}n_{\nu}=0,
\end{eqnarray}
which are indeed satisfied by Eqs.~(\ref{Gmunu_R_exp}) and (\ref{Gmunu_L_exp}).

The real and imaginary parts of Eq.~(\ref{EOM_WT}) read 
\begin{eqnarray}\nonumber
&&\frac{\hbar^2}{4}(\eta^{\mu}_{\mbox{ }\rho}\partial^2-\partial_{\rho}\partial^{\mu})G^{h<\rho\nu}_{\rm S}
-(\eta^{\mu}_{\mbox{ }\rho}q^2-q_{\rho}q^{\mu})G^{h<\rho\nu}_{\rm S}+\frac{\hbar}{2}(2\eta^{\mu}_{\mbox{ }\rho}q\cdot\partial-q_{\rho}\partial^{\mu}-q^{\mu}\partial_{\rho})G^{h<\rho\nu}_{\rm A}=0\,,
\\
&&\frac{\hbar^2}{4}(\eta^{\mu}_{\mbox{ }\rho}\partial^2-\partial_{\rho}\partial^{\mu})G^{h<\rho\nu}_{\rm A}
-(\eta^{\mu}_{\mbox{ }\rho}q^2-q_{\rho}q^{\mu})G^{h<\rho\nu}_{\rm A}-\frac{\hbar}{2}(2\eta^{\mu}_{\mbox{ }\rho}q\cdot\partial-q_{\rho}\partial^{\mu}-q^{\mu}\partial_{\rho})G^{h<\rho\nu}_{\rm S}=0\,,
\end{eqnarray}
which further reduce to 
\begin{eqnarray}\nonumber
&&\Big(\frac{\hbar^2}{4}\partial^2-q^2\Big)G_{\rm S}^{h<\mu\nu}+\hbar q\cdot\partial G_{\rm A}^{h<\mu\nu}=0\,,
\\
&&\Big(\frac{\hbar^2}{4}\partial^2-q^2\Big)G_{\rm A}^{h<\mu\nu}-\hbar q\cdot\partial G_{\rm S}^{h<\mu\nu}=0\,,
\end{eqnarray}
by using the constraints from the Coulomb gauge in Eq.~(\ref{constraint_Coulomb}). Due to the symmetric and anti-symmetric properties of $G^{h<\mu\nu}_{\rm S}$ and $G^{h<\mu\nu}_{\rm A}$, we thus find the master equations dictating the dynamics of $G^{h<\mu\nu}_{\rm S/A}$ read
\begin{eqnarray}\label{Meq_on-shell}
&&\Big(q^2-\frac{\hbar^2}{4}\partial^2\Big)G_{\rm S/A}^{h<\mu\nu}=0\,,
\\\label{Meq_KT}
&&q\cdot\partial G_{\rm S/A}^{h<\mu\nu}=0\,,
\end{eqnarray}
where Eq.~(\ref{Meq_on-shell}) governs the on-shell condition and Eq.~(\ref{Meq_KT}) gives the free-streaming kinetic equation. On the other hand, Eq.~(\ref{constraint_Coulomb}) can be utilized to solve for the Wigner functions of photons. However, Eq.~(\ref{Meq_KT}) cannot uniquely determine the $\hbar$ corrections of photons. When taking a constant frame vector $n^{\mu}$, Eq.~(\ref{Meq_on-shell}) results in $\delta(q^2)$ term in Wigner functions up to $\mathcal{O}(\hbar)$ and Eq.~(\ref{Meq_KT}) simply leads to just the ordinary Boltzmann equation $\delta(q^2)q\cdot\partial f^{\gamma}_h=0$ or equivalently $\delta(q^2)q\cdot\partial f^{\gamma}_{\rm V/A}=0$. In such a case, the free-streaming part is unmodified by the $\hbar$ corrections for polarized photons analogous to the case for fermions in the absence of background fields, whereas the collision term is more involved. 

\subsection{Collisions}
We may follow the standard approach based on the Dyson-Schwinger equation and the real-time formalism to systematically incorporate the collision term for the kinetic theory of photons in light of a similar derivation for fermions \cite{Blaizot:2001nr,Hidaka:2016yjf}.
However, such a derivation is technically more involved due to the tensor structure of photon Wigner functions as shown in Appendix~\ref{app_photon_KT}. Because of the axial part of Wigner functions and the involvement of $\hbar$ corrections, the derivation of QKT for photons is nontrivial as opposed to the derivation for the classical kinetic theory in Ref.~\cite{Blaizot:1999xk}. Eventually, for the full photon propagator, we obtain a master equation up to $\mathcal{O}(\hbar^3)$
\begin{eqnarray}\label{general_KT_col}\nonumber
&&\Big(q^2-\frac{\hbar^2}{4}\partial^2+{\rm i}\hbar q\cdot\partial\Big)G^{<\mu\nu}
+\hbar {\cal P}^{\mu\rho}\Big(
{\selfEnergy}^{+}_{\rho\sigma}\star G^{<\sigma\nu}+{\selfEnergy}^{<}_{\rho\sigma}\star G^{+\sigma\nu}\Big)\notag\\
&&=\frac{\ri\hbar}{2} {\cal P}^{\mu\rho}\big({\selfEnergy}^>_{\rho\sigma}\star G^{<\sigma\nu}-{\selfEnergy}^<_{\rho\sigma}\star G^{>\sigma\nu}\big)\,,
\end{eqnarray}
where $(\widehat{A B})_{\rho}^{\,\,\nu}= A^{>}_{\rho\sigma}B^{<\sigma\nu}-A^{<}_{\rho\sigma}B^{>\sigma\nu}$ and $A\star B\equiv AB+\frac{{\rm i}\hbar}{2}A*B+{\cal O}(\hbar^2)$ with
$A*B\equiv (\partial_{q_{\alpha}}A)(\partial_{\alpha}B)-(\partial_{\alpha}A)(\partial_{q_{\alpha}}B)$ and 
\begin{eqnarray}
{\cal P}^{\mu\nu}(q)= P^{\mu\nu}(q)-\frac{{\rm i}\hbar}{2}\delta P^{\mu\nu}(q)\,,
\qquad
\delta P^{\mu\nu}(q)=\frac{1}{|\bm q|^2}q^{(\mu}_{\perp}\Big(\partial^{\nu)}_{\perp}+\hat{q}^{\nu)}_{\perp}\hat{q}_{\perp}\cdot\partial_{\perp}\Big)\,.
\end{eqnarray}
This master equation then gives rise to the kinetic equations and on-shell constraint equations up to $\mathcal{O}(\hbar)$. 
Here ${\selfEnergy}^{\lessgtr}_{\mu\nu}$ denote the lesser/greater self-energies of photons. We also introduced the retarded and advanced propagators in Eq.~(\ref{R_A}) with a similar definition applied to the retarded and advanced self-energies for photons. Moreover, we define $\selfEnergy^+_{\sigma\rho}\equiv (\selfEnergy^{\rm ret}_{\sigma\rho}+\selfEnergy^{\rm adv}_{\sigma\rho})/2$ and $G^{+}_{\sigma\rho}\equiv {(G^{\rm ret}_{\sigma\rho}+G^{\rm adv}_{\sigma\rho})/2}$. In light of the derivation for CKT of fermions, we will take $\selfEnergy^+_{\sigma\rho}=G^{+}_{\sigma\rho}=0$, which corresponds to the vanishing real part of the retarded self-energy and of the retarded propagator for the fermionic case. 
The $\hbar$ terms in Eq.~(\ref{general_KT_col}) are essential to satisfy the gauge constraint (\ref{constraint}).

We may decompose the self-energies into real and imaginary parts, ${\selfEnergy}^{\lessgtr}_{\mu\nu}=({\selfEnergy}^{\lessgtr}_{\text{Re}})_{\mu\nu}+{\rm i}({\selfEnergy}^{\lessgtr}_{\text{Im}})_{\mu\nu}$. Given that ${\selfEnergy}^{\lessgtr}_{\mu\nu}$ are Hermitian, we find $({\selfEnergy}^{\lessgtr}_{\text{Re}})_{\mu\nu}=({\selfEnergy}^{\lessgtr}_{\text{Re}})_{\nu\mu}$ and $({\selfEnergy}^{\lessgtr}_{\text{Im}})_{\mu\nu}=-({\selfEnergy}^{\lessgtr}_{\text{Im}})_{\nu\mu}$. Decomposing Eq.~(\ref{general_KT_col}) into symmetric and anti-symmetric parts and taking $\selfEnergy^+_{\sigma\rho}=G^{+}_{\sigma\rho}=0$, we derive the kinetic equations [see Eqs. (\ref{cons_eq_full_1})-(\ref{kin_eq_full_S})],
\begin{eqnarray}\nonumber
q\cdot\partial G^{<\mu\nu}_{\rm S}&=&-\frac{1}{4}P^{(\mu\rho}\Big[(\widehat{{\selfEnergy}_{\text{Re}}G_{\rm S}})_{\rho}^{\,\,\nu)}-(\widehat{{\selfEnergy}_{\text{Im}}G_{\rm A}})_{\rho}^{\,\,\nu)}\Big]
-\frac{\hbar}{8}\delta P^{(\mu\rho}(q)\Big[(\widehat{{\selfEnergy}_{\text{Re}}G_{\rm A}})_{\rho}^{\,\,\nu)}+(\widehat{{\selfEnergy}_{\text{Im}}G_{\rm S}})_{\rho}^{\,\,\nu)}\Big]
\\\nonumber
&&+\frac{\hbar}{8}P^{[\mu\rho}\Big[(\widehat{{\selfEnergy}_{\text{Re}}* G_{\rm A}})_{\rho}^{\,\,\nu]}+(\widehat{{\selfEnergy}_{\text{Im}}*G_{\rm S}})_{\rho}^{\,\,\nu]}\Big]
\,,
\\\nonumber
q\cdot\partial G^{<\mu\nu}_{\rm A}&=&-\frac{1}{4}P^{[\mu\rho}\Big[(\widehat{{\selfEnergy}_{\text{Re}}G_{\rm A}})_{\rho}^{\,\,\nu]}+(\widehat{{\selfEnergy}_{\text{Im}}G_{\rm S}})_{\rho}^{\,\,\nu]}\Big]
+\frac{\hbar}{8}\delta P^{[\mu\rho}(q)
\Big[(\widehat{{\selfEnergy}_{\text{Re}}G_{\rm S}})_{\rho}^{\,\,\nu]}-(\widehat{{\selfEnergy}_{\text{Im}}G_{\rm A}})_{\rho}^{\,\,\nu]}\Big]
\\
&&-\frac{\hbar}{8}P^{[\mu\rho}\Big[(\widehat{{\selfEnergy}_{\text{Re}}* G_{\rm S}})_{\rho}^{\,\,\nu]}-(\widehat{{\selfEnergy}_{\text{Im}}* G_{\rm A}})_{\rho}^{\,\,\nu]}\Big]
\,,
\end{eqnarray}
and the constraint equations dictating the on-shell conditions 
\begin{eqnarray}\nonumber
q^2G^{<\mu\nu}_{\rm S}&=&\frac{\hbar}{4} P^{(\mu\rho}\Big[(\widehat{{\selfEnergy}_{\text{Re}} G_{\rm A}})_{\rho}^{\,\,\nu)}+(\widehat{{\selfEnergy}_{\text{Im}} G_{\rm S}})_{\rho}^{\,\,\nu)}
\Big]\,,
\\
q^2G^{<\mu\nu}_{\rm A}&=&-\frac{\hbar}{4} P^{[\mu\rho}\Big[(\widehat{{\selfEnergy}_{\text{Re}} G_{\rm S}})_{\rho}^{\,\,\nu]}-(\widehat{{\selfEnergy}_{\text{Im}} G_{\rm A}})_{\rho}^{\,\,\nu]}
\Big]\,,
\end{eqnarray} 
up to $\mathcal{O}(\hbar)$.

\subsection{Effective QKT for photons}
As mentioned in Sec.~\ref{sec_CS_current}, we may further apply the power-counting scheme such that ${f}_{\rm V}^{\gamma} = \mathcal{O}(\hbar^0)$ and ${f}_{\rm A}^{\gamma} = \mathcal{O}(\hbar)$, which hence yield $G^{<\mu\nu}_{\rm S} = \mathcal{O}(\hbar^0)$ and $G^{<\mu\nu}_{\rm A} = \mathcal{O}(\hbar)$. We also assume $({\selfEnergy}^{\lessgtr}_{\text{Re}})^{\mu}_{\rho} = \mathcal{O}(\hbar^0)$ and $({\selfEnergy}^{\lessgtr}_{\text{Im}})^{\mu}_{\rho} = \mathcal{O}(\hbar)$. The effective kinetic theory then becomes 
\begin{eqnarray}\nonumber\nonumber\label{KE_photons}
q\cdot\partial G^{<\mu\nu}_{\rm S}&=&-\frac{1}{4}P^{(\mu\rho}(\widehat{{\selfEnergy}_{\text{Re}}G_{\rm S}})_{\rho}^{\,\,\nu)}\,,
\\\nonumber
q\cdot\partial G^{<\mu\nu}_{\rm A}&=&-\frac{1}{4}P^{[\mu\rho}\Big[(\widehat{{\selfEnergy}_{\text{Re}}G_{\rm A}})_{\rho}^{\,\,\nu]}+(\widehat{{\selfEnergy}_{\text{Im}}G_{\rm S}})_{\rho}^{\,\,\nu]}\Big]
+\frac{\hbar}{8}\delta P^{[\mu\rho}(q)(\widehat{{\selfEnergy}_{\text{Re}}G^{(f)}_{\rm S}})_{\rho}^{\,\,\nu]}
-\frac{\hbar}{8}P^{[\mu\rho}(\reallywidehat{{\selfEnergy}_{\text{Re}}* G_{\rm S}})_{\rho}^{\,\,\nu]}\,,
\\
\end{eqnarray}
with the constraint equations 
\begin{eqnarray}
q^2 G^{<\mu\nu}_{\rm S}=q^2 G^{<\mu\nu}_{\rm A}=0.
\end{eqnarray}
Note that contracting the first equation in Eq.~(\ref{KE_photons}) with $-\eta_{\mu\nu}/2$ leads to the ordinary Boltzmann equation for photons,
\begin{eqnarray}\label{Boltzmann_eq}
q\cdot\partial {f}_{\rm V}^{\gamma}=\frac{1}{4}P^{\mu\rho}(\widehat{{\selfEnergy}_{\text{Re}}\hat{G}_{\rm S}})_{\rho\mu}
=-\frac{1}{4}{\selfEnergy}^{>\mu\rho}_{\text{Re}} \hat{G}^{<}_{{\rm S}\rho\mu}+\frac{1}{4}{\selfEnergy}^{<\mu\rho}_{\text{Re}} \hat{G}^{>}_{{\rm S}\rho\mu}\,,
\end{eqnarray}
where we introduced the shorthand notations
\begin{eqnarray}
\hat{G}^{<\mu\nu}_{\rm S/A} \equiv \frac{G^{<\mu\nu}_{\rm S/A}}{2\pi\delta(q^2)\mathrm{sgn}(q\cdot n)}\,.
\end{eqnarray}
From Eq.~(\ref{Boltzmann_eq}), one can read out the self-energies by comparing the collision term above with scattering cross section. 

Since 
\begin{eqnarray}\label{input_GA_free}
q\cdot\partial \hat{G}^{<\mu\nu}_{\rm A}=-q\cdot\partial\left(S_{\gamma}^{\mu\nu}{f}_{\rm A}^{\gamma}+\frac{\hbar}{2|\bm q|}\hat{q}^{[\mu}_{\perp}\partial^{\nu]}_{\perp}{f}_{\rm V}^{\gamma}\right)
=-S_{\gamma}^{\mu\nu}q\cdot\partial{f}_{\rm A}^{\gamma}+\frac{\hbar}{8|\bm q|}\hat{q}^{[\mu}_{\perp}\partial^{\nu]}_{\perp}(\widehat{{\selfEnergy}_{\text{Re}}\hat{G}_{\rm S}})^{\rho}_{\,\,\rho}
\end{eqnarray}
by using Eq.~(\ref{Boltzmann_eq}), we may rewrite the kinetic equation for $G^{<\mu\nu}_{\rm A}$ as
\begin{eqnarray}\nonumber\label{KE_fA_temp}
0&=&\delta(k^2)\Bigg(S_{\gamma}^{\mu\nu}q\cdot\partial{f}_{\rm A}^{\gamma}-\frac{1}{4}P^{[\mu\rho}\Big[(\widehat{{\selfEnergy}_{\text{Re}}\hat{G}_{\rm A}})_{\rho}^{\,\,\nu]}+(\widehat{{\selfEnergy}_{\text{Im}}\hat{G}^{(f)}_{\rm S}})_{\rho}^{\,\,\nu]}\Big]
+\frac{\hbar}{8|\bm q|}\Big[|\bm q|\delta P^{[\mu\rho}(\widehat{{\selfEnergy}_{\text{Re}}\hat{G}_{\rm S}})_{\rho}^{\,\,\nu]}
\\
&& \qquad \quad -\hat{q}^{[\mu}_{\perp}\partial^{\nu]}_{\perp}(\widehat{{\selfEnergy}_{\text{Re}}\hat{G}_{\rm S}})^{\rho}_{\,\,\rho}\Big]
-\frac{\hbar}{8}P^{[\mu\rho}(\reallywidehat{{\selfEnergy}_{\text{Re}} * \hat{G}_{\rm S}})_{\rho}^{\,\,\nu]}\Bigg)
+\frac{\hbar}{4}P^{[\mu\rho}\big[\reallywidehat{(q\cdot\partial{\selfEnergy}_{\text{Re}}) \hat{G}_{\rm S}}\big]_{\rho}^{\,\,\nu]}\delta'(q^2)\,.
\end{eqnarray}
One may wonder whether the possible corrections to the tensor structure of $\hat{G}_{\rm A}^{<\mu\nu}$ from collisions should be included in Eq.~(\ref{input_GA_free}). Nonetheless, such corrections will be at higher orders in our power counting. Due to the gauge constraint (\ref{constraint}), the right-hand side of Eq.~(\ref{KE_fA_temp}) should be transverse to both $n^{\mu}$ and $q_{\perp}^{\mu}$. Except for $\epsilon^{\mu\nu\alpha\beta}q_{\alpha}n_{\beta}$, the only possible tensor structure is given by $\mathcal{A}^{[\mu}\mathcal{B}^{\nu]}$, where $\mathcal{A}^{\mu}$ and $\mathcal{B}^{\mu}$ are two new timelike vectors characterizing anisotropy of systems such that $\mathcal{A}\cdot n=\mathcal{A}\cdot q_{\perp}=0$ and so does $\mathcal{B}^{\mu}$. In addition, $\mathcal{A}^{\mu}$ and $\mathcal{B}^{\mu}$ should accordingly serve as the source for non-vanishing ${f}_{\rm A}^{\gamma}$. However, according to our power counting, $\mathcal{A}^{\mu}$ and $\mathcal{B}^{\mu}$ have to be led by gradient terms and at $\mathcal{O}(\hbar)$. Consequently, we conclude that $\mathcal{A}^{[\mu}\mathcal{B}^{\nu]} = \mathcal{O}(\hbar^2)$ and the right-hand side of Eq.~(\ref{KE_fA_temp}) is proportional to $S_{\gamma}^{\mu\nu}$ for systems we considered.  

In practice, it is sometime more convenient to rewrite 
\begin{eqnarray}\label{rewrite_star}
-\frac{\hbar}{8}P^{[\mu\rho}(\reallywidehat{{\selfEnergy}_{\text{Re}} * \hat{G}_{\rm S}})_{\rho}^{\,\,\nu]}
=-\frac{\hbar}{8}P^{[\mu\rho}\Big[\partial_{q_{\alpha}}(\reallywidehat{{\selfEnergy}_{\text{Re}} \partial_{\alpha} \hat{G}_{\rm S}})_{\rho}^{\,\,\nu]}
-\partial_{\alpha}(\reallywidehat{{\selfEnergy}_{\text{Re}} \partial_{q_{\alpha}} \hat{G}_{\rm S}})_{\rho}^{\,\,\nu]}\Big]\,.
\end{eqnarray}
Also, we may analyze the last term involving $\delta'(q^2)$ in Eq.~(\ref{KE_fA_temp}). This term can be more explicitly written as
\begin{eqnarray}
\hbar P^{[\mu\rho}\big[\reallywidehat{(q\cdot\partial{\selfEnergy}_{\text{Re}}) \hat{G}_{\rm S}}\big]_{\rho}^{\,\,\nu]}\delta'(q^2)
=\hbar\delta'(q^2)P^{[\mu\rho} P^{\sigma\nu]}\Big[(q\cdot\partial{\selfEnergy}^>_{\text{Re}\rho\sigma}){f}_{\rm V}^{\gamma}
-(q\cdot\partial{\selfEnergy}^<_{\text{Re}\rho\sigma})(1+{f}_{\rm V}^{\gamma})
\Big]
\end{eqnarray}
up to $\mathcal{O}(\hbar)$.
Although $(q\cdot\partial{\selfEnergy}^{\gtrless}_{\text{Re}\rho\sigma})$ above could be off-shell, both $P^{\mu\rho}$ and $P^{\sigma\nu}$ therein are on-shell. Provided $(q\cdot\partial{\selfEnergy}^{\gtrless}_{\text{Re}\rho\sigma})$ are symmetric as their generic property, we find $\hbar P^{[\mu\rho}\big[\reallywidehat{(q\cdot\partial{\selfEnergy}_{\text{Re}}) \hat{G}_{\rm S}}\big]_{\rho}^{\,\,\nu]}\delta'(q^2)=0$. 
Then the effective QKT for on-shell photons reduces to 
\begin{eqnarray}\nonumber\label{KE_GA}
S_{\gamma}^{\mu\nu}q\cdot\partial{f}_{\rm A}^{\gamma}&=&\frac{1}{4}P^{[\mu\rho}\Big[(\widehat{{\selfEnergy}_{\text{Re}}\hat{G}_{\rm A}})_{\rho}^{\,\,\nu]}+(\widehat{{\selfEnergy}_{\text{Im}}\hat{G}_{\rm S}})_{\rho}^{\,\,\nu]}\Big]
-\frac{\hbar}{8|\bm q|}\Big[|\bm q|\delta P^{[\mu\rho}(\widehat{{\selfEnergy}_{\text{Re}}\hat{G}_{\rm S}})_{\rho}^{\,\,\nu]}
-\hat{q}^{[\mu}_{\perp}\partial^{\nu]}_{\perp}(\widehat{{\selfEnergy}_{\text{Re}}\hat{G}_{\rm S}})^{\rho}_{\,\,\rho}\Big]
\\
&&
+\frac{\hbar}{8}P^{[\mu\rho}\Big[\partial_{q_{\alpha}}(\reallywidehat{{\selfEnergy}_{\text{Re}} \partial_{\alpha} \hat{G}_{\rm S}})_{\rho}^{\,\,\nu]}
-\partial_{\alpha}(\reallywidehat{{\selfEnergy}_{\text{Re}} \partial_{q_{\alpha}} \hat{G}_{\rm S}})_{\rho}^{\,\,\nu]}\Big]\,.
\end{eqnarray} 
Finally, we may rewrite Eq.~(\ref{KE_GA}) as a scalar kinetic equation,
\begin{eqnarray}\nonumber\label{KE_fA}
q\cdot\partial{f}_{\rm A}^{\gamma}&=&\frac{1}{4}S^{\gamma}_{\mu\nu}\Bigg(P^{\mu\rho}\Big[(\widehat{{\selfEnergy}_{\text{Re}}\hat{G}_{\rm A}})_{\rho}^{\,\,\nu}+(\widehat{{\selfEnergy}_{\text{Im}}\hat{G}_{\rm S}})_{\rho}^{\,\,\nu}\Big]
-\frac{\hbar}{2|\bm q|}\Big[|\bm q|\delta P^{\mu\rho}(\widehat{{\selfEnergy}_{\text{Re}}\hat{G}_{\rm S}})_{\rho}^{\,\,\nu}
-\hat{q}^{\mu}_{\perp}\partial^{\nu}_{\perp}(\widehat{{\selfEnergy}_{\text{Re}}\hat{G}_{\rm S}})^{\rho}_{\,\,\rho}\Big]
\\
&&
+\frac{\hbar}{2}P^{\mu\rho}\Big[\partial_{q_{\alpha}}(\reallywidehat{{\selfEnergy}_{\text{Re}} \partial_{\alpha} \hat{G}_{\rm S}})_{\rho}^{\,\,\nu}
-\partial_{\alpha}(\reallywidehat{{\selfEnergy}_{\text{Re}} \partial_{q_{\alpha}} \hat{G}_{\rm S}})_{\rho}^{\,\,\nu}\Big]\Bigg)\,,
\end{eqnarray} 
where we used $S^{\gamma}_{\mu\nu}S_{\gamma}^{\mu\nu}=2$ for on-shell photons. Eventually, Eqs.~(\ref{Boltzmann_eq}) and (\ref{KE_fA}) jointly delineate the evolution of ${f}^{\gamma}_{{\rm V/A}}$ in phase space. As mentioned in the beginning of this section, in practice, one could read out the structure of self-energies through the collision term in Eq.~(\ref{Boltzmann_eq}) as the standard Boltzmann equation. Inputting the Wigner functions with $\hbar$ corrections, one is able to evaluate the collision term in Eq.~(\ref{KE_fA}), which systematically captures the quantum corrections.

\section{Summary and Outlook}\label{sec_sum_outlook}
In this paper, we have derived the Wigner functions and QKT of polarized photons in the Coulomb gauge up to $\mathcal{O}(\hbar)$ based on QED. We found that the Wigner functions incorporate anti-symmetric and imaginary components characterizing the helicity distribution in phase space, which are also responsible for the CS and zilch currents. In particular, the derivation analogous to the fermionic case reveals the absorption of Berry connections into distribution functions of polarized photons, which hence manifests itself in the frame-dependence of the distribution functions. We also discussed the photonic CVE and ZVE triggered by fluid vorticity and our findings are in agreements with some of previous studies. This QKT enables us to track both the number (or energy) and spin densities of photons (dictated by ${f}_{\rm V}^{\gamma}$ and ${f}_{\rm A}^{\gamma}$ more precisely) with collisions in terms of the self-energies. Adopting suitable power counting, the QKT boils down to simplified scalar kinetic equations for practical applications.   

There are several future directions. First, combined with the effective QKT for fermions recently developed in Ref.~\cite{Yang:2020hri}, we can further investigate the intertwined spin transport between electrons and photons in QED.%
\footnote{Notably, the formalism in Ref.~\cite{Yang:2020hri} has been lately applied to the Nambu--Jona-Lasinio model at the quark level, which explicitly shows the spin polarization triggered by vorticity corrections upon the collision term from detailed balance \cite{Wang:2020pej}.}
In particular, it is curious whether the ZVE derived by demanding the explicit frame independence in thermal equilibrium in the present study remains unchanged when considering the helicity transfer from electrons through collisions. Second, similarly to the case of massless Dirac fermions \cite{Liu:2018xip}, it would be interesting to extend our formalism in curved spacetime, where an effective refractive index should lead to the photonic spin Hall effect. Third, since the present formalism of photons can be directly applied to weakly coupled gluons in the absence of background color fields, we may study a similar scenario for entangled spin transport between quarks and gluons in QCD. Although more complicated scattering processes are involved, this direction would be crucial to understand the dynamical spin polarization pertinent to heavy-ion experiments. We may also generalize the present formalism to the gluonic case with the inclusion of background color fields, which could have potential applications to the chirality transfer in QCD at the early stage of the relativistic heavy-ion collisions. 
Finally, our formalism may also be applied to other non-equilibrium systems involving polarized photons in condensed matter physics and astrophysics.

\acknowledgments
This work was supported by JSPS KAKENHI Grant Numbers~17H06462, 18H01211, 19K03852, 20K14470, and 20K03948. 
N.~Y. and D.-L. Y. were supported by Keio Institute of Pure and Applied Sciences (KiPAS) project in Keio University.

\appendix
\section{Derivation of $\Pi_{\mu\nu}^{(0)}$ and $\Pi_{\mu\nu\alpha}^{(1)}$}
\label{app_Pi}
In this Appendix, we provide the detail of the derivations of $\Pi_{\mu\nu}^{(0)}$ and $\Pi_{\mu\nu\alpha}^{(1)}$ in Eqs.~(\ref{Pimunu0}) and (\ref{Pimunu1}). 
Let us first compute $\Pi_{\mu\nu}^{(0)}(q)$. Notice the relations,
\begin{eqnarray}
\label{eq:ccdagger}
c_{\rm R}^{(+)}c_{\rm R}^{(+)\dagger}=\frac{1}{2|{\bm q}|}\bar{\sigma}^{\nu}q_{\nu}\,,
\qquad
c_{\rm R}^{(-)}c_{\rm R}^{(-)\dagger}=\frac{1}{2|{\bm q}|}\bar{\sigma}^{\nu}\bar{q}_{\nu}\,,
\end{eqnarray}
where $\bar{q}^{\mu}\equiv(q\cdot n)n^{\mu}-q_{\perp}^{\mu}$. 
Therefore, we have 
\begin{eqnarray}
\Pi_{\mu\nu}^{(0)}(q)=\frac{1}{8|{\bm q}|^2} \text{tr}\big(\bar{\sigma}^{\alpha}\sigma_{\mu}\bar{\sigma}^{\beta}\sigma_{\nu}\big) q_{\alpha}\bar{q}_{\beta}\,.
\end{eqnarray}
Then, we obtain Eq.~(\ref{Pimunu0}) by using $q\cdot n=|{\bm q}|$ and the result of the trace%
\footnote{This relation follows from the traces in the four-component forms,
$ \tr[\gam^\a \gam^\mu \gam^\b \gam^\nu ] $ and $ \tr[\gam^5\gam^\a \gam^\mu \gam^\b \gam^\nu ]$.}
\begin{eqnarray}
\kappa^{\alpha \mu \beta \nu} \equiv \frac{1}{2}
\text{tr}\left(\bar{\sigma}^{\alpha}\sigma^{\mu}\bar{\sigma}^{\beta}\sigma^{\nu}\right)=\eta^{\alpha(\mu}\eta^{\nu)\beta}-\eta^{\alpha\beta}\eta^{\mu\nu}-{\rm i}\epsilon^{\alpha\mu\beta\nu}\,.
\label{eq:trace}
\end{eqnarray}

Next, we compute $\Pi_{\mu\nu\alpha}^{(1)}(q)$. 
According to Eq.~(\ref{eq:ccdagger}) and 
an identity $\sigma^{\nu}\bar{\sigma}^{\alpha}\sigma^{\mu}=\kappa^{\nu\alpha\mu\beta}\sigma_{\beta}$, we have  
\begin{eqnarray}\nonumber
\Pi_{\mu\nu\alpha}^{(1)}(q)&=&\frac{\rm i}{2|{\bm q}|}\text{tr}\Big[-{\cal A}\Big(\partial_{q^{\alpha}}c_{\rm R}^{(-)}c_{\rm R}^{(-)\dagger}\Big)
\sigma_{\mu}\bar{\sigma}_{\rho}\sigma_{\nu}q^{\rho}
+\sigma_{\nu}\bar{\sigma}_{\rho}\sigma_{\mu}\bar{q}^{\rho}{\cal A}\Big(\partial_{q^{\alpha}}c_{\rm R}^{(+)}c_{\rm R}^{(+)\dagger}\Big)\Big]
\\
&=&\frac{\rm i}{2|{\bm q}|}\text{tr}\Big[-{\cal A}\Big(\partial_{q^{\alpha}}c_{\rm R}^{(-)}c_{\rm R}^{(-)\dagger}\Big)
\sigma^{\beta}q^{\rho}\kappa_{\mu\rho\nu\beta}
+{\cal A}\Big(\partial_{q^{\alpha}}c_{\rm R}^{(+)}c_{\rm R}^{(+)\dagger}\Big)\sigma^{\beta}\bar{q}^{\rho}\kappa_{\nu\rho\mu\beta}\Big]\,.
\end{eqnarray}
Then, we use 
\begin{eqnarray}
\text{tr}\Big[{\cal A}\Big( \partial_{q^{\alpha}}c_{\rm R}^{(\pm)}c_{\rm R}^{(\pm)\dagger}\Big)\sigma^{\beta}\Big]
= \text{tr}\Big[{\cal A}\Big( \partial_{q^{\alpha}}c_{\rm R}^{(\pm)}c_{\rm R}^{(\pm)\dagger}\sigma^{\beta} \Big)\Big]
= {\cal A}\Big(c_{\rm R}^{(\pm)\dagger}\sigma^{\beta}\partial_{q^{\alpha}}c_{\rm R}^{(\pm)}\Big)
\, ,
\end{eqnarray}
and
\begin{subequations}
\label{Im_cDc_+}
\begin{eqnarray}
{\cal A}\left( c^{(+)\dagger}_{\rm R}\sigma^{\beta}\partial_{q_{\alpha}} c^{(+)}_{\rm R}\right)
&=&-a^{\alpha}_{+}(q)\frac{q^{\beta}}{|{\bm q}|}-\frac{1}{2|{\bm q}|}n_{\lambda}\epsilon^{\lambda\beta\sigma\alpha}\hat{q}_{\sigma}\,,
\\
{\cal A}\left( c^{(-)\dagger}_{\rm R}\sigma^{\beta} \partial_{q_{\alpha}} c^{(-)}_{\rm R}\right)
&=&-a^{\alpha}_{-}(q)\frac{\bar{q}^{\beta}}{|{\bm q}|}-\frac{1}{2|{\bm q}|}n_{\lambda}\epsilon^{\lambda\beta\sigma\alpha}\hat{q}_{\sigma}\,,
\end{eqnarray}
\end{subequations}
where $a^{\mu}_{\pm}(q)$ are the Berry connections given in Eq.~(\ref{Berry}). 
Plugging those expressions, we obtain
\begin{eqnarray}\nonumber
\Pi_{\mu\nu\alpha}^{(1)}(q)&=&\frac{\rm i}{2|{\bm q}|^2}\Big[\Big(a_{-\alpha}(q)\bar{q}^{\beta}+\frac{1}{2}n_{\lambda}\epsilon^{\lambda\beta\sigma\gamma}\eta_{\gamma\alpha}\hat{q}_{\sigma}\Big)q^{\rho}\kappa_{\mu\rho\nu\beta}
-\Big(a_{+\alpha}(q)q^{\beta}+\frac{1}{2}n_{\lambda}\epsilon^{\lambda\beta\sigma\gamma}\eta_{\gamma\alpha}\hat{q}_{\sigma}\Big)\bar{q}^{\rho}\kappa_{\nu\rho\mu\beta}
\Big]
\, .
\end{eqnarray}
Inserting $ \kappa^{\alpha \mu \beta \nu}  $ given above, this can be further written as Eq.~(\ref{Pimunu1}).

\section{Energy-momentum tensor of photons}
\label{app_EM_tensor}
We consider the Belinfante energy-momentum (EM) tensor of photons, 
\begin{eqnarray}
T^{\mu\nu}= - F^{\mu\rho}F_{\ \, \rho}^{\nu}+\frac{\eta^{\mu\nu}}{4} F^{\alpha\beta}F_{\alpha\beta}
= - \frac{1}{2}F^{(\mu}_{\ \ \, \rho}F^{\nu)\rho} + \frac{\eta^{\mu\nu}}{4} F^{\alpha\beta}F_{\alpha\beta}
\, ,
\end{eqnarray}
which is a traceless and symmetric tensor. 
The first term can be written as
\begin{eqnarray}\nonumber
F^{(\mu}_{\ \ \, \rho}F^{\nu)\rho} (x)
&=&
\partial^{(\mu}_{y}\partial^{\nu)}_{x}A^{\rho}(y)A_{\rho}(x)
+ \partial^{\rho}_{y}\partial_{x\rho}A^{(\mu}(y)A^{\nu)}(x)
\nn
\\
&&
-\partial^{(\mu}_{y}\partial_{x\rho}A^{\rho}(y)A^{\nu)}(x)
-\partial^{(\mu}_{x}\partial_{y\rho}A^{\nu)}(y)A^{\rho}(x)
\, ,
\end{eqnarray}
which yields the following form after the Wigner transformation: 
\begin{eqnarray}\nonumber
\langle F^{(\mu}_{\ \ \, \rho}F^{\nu)\rho} \rangle (q,X)
&=&
- \hat{\Delta}^{(\mu}_{\quad\rho}G^{<\nu)\rho}(q,X)
- \hat{\Delta}^{\,\,(\mu}_{\rho}G^{<\rho\nu)}(q,X)
+ \hat{\Delta}^{(\mu\nu)}G^{<\rho}_{\,\,\rho}(q,X)
+ \hat{\Delta}^{\,\,\rho}_{\rho}G^{<(\nu\mu)}(q,X)
\\\nonumber
&=&
- \frac{1}{\hbar^2}\Big[q_{\rho}q^{(\mu}\Big(G^{<\nu)\rho}+G^{<\rho\nu)}\Big)
+\frac{{\rm i}\hbar}{2}\big(q^{(\mu}\partial_{\rho}-q_{\rho}\partial^{(\mu}\big)\Big(G^{<\nu)\rho}-G^{<\rho\nu)}\Big)
\\
&&-q^{(\mu}q^{\nu)}G^{<\rho}_{\,\,\rho}
-q^2G^{<(\nu\mu)}+\mathcal{O}(\hbar^2)
\Big]
\,.
\end{eqnarray}
Here, we defined a derivative operator  
\begin{eqnarray}
\hat{\Delta}_{\mu\rho}=
\frac{1}{\hbar^2}\Big(q_{\mu}-\frac{{\rm i}\hbar}{2}\partial_{\mu}\Big)\Big(q_{\rho}+\frac{{\rm i}\hbar}{2}\partial_{\rho}\Big)
=\frac{q_{\mu}q_{\rho}}{\hbar^2}+\frac{\rm i}{2\hbar}q_{[\mu}\partial_{\rho]}+\frac{1}{4}\partial_{\mu}\partial_{\rho}\,.
\end{eqnarray}
By further implementing the Coulomb-gauge constraint~(\ref{constraint}), 
we find
\begin{eqnarray}
\langle F^{(\mu}_{\ \ \, \rho}F^{\nu)\rho} \rangle (q,X) = 
- \frac{1}{\hbar^2}\Big[{\rm i}\hbar q^{(\mu}\partial_{\perp\rho}\Big(G^{<\nu)\rho}-G^{<\rho\nu)}\Big)
-q^{(\mu}q^{\nu)}G^{<\rho}_{\,\,\rho}
-q^2G^{<(\nu\mu)}+\mathcal{O}(\hbar^2)
\Big]
\,.
\end{eqnarray}
Inserting the lesser propagator (\ref{Gmunu_full}), we find 
\begin{eqnarray}
\hbar^2\langle F^{(\mu}_{\ \ \, \rho}F^{\nu)\rho}   \rangle (q,X) =
-4\pi\sgn(q_0)\delta(q^2)\Big(
2q^{\mu}q^{\nu}{f}_{\rm V}^{\gamma}
+\hbar q^{(\mu}S_{\gamma}^{\nu)\rho}\partial_{\rho}{f}_{\rm A}^{\gamma}
\Big)
+\mathcal{O}(\hbar^2)
\,,
\end{eqnarray}
where $q_0=q\cdot n$. Recall that we took $n^{\mu}=(1,\bm 0)$.
Accordingly, one finds $\hbar^2\langle F^{\alpha\beta} F_{\alpha\beta} \rangle (q,X)=0$, 
and thus the phase-space electromagnetic energy density is given by 
\begin{eqnarray}
\langle T^{\mu\nu}\rangle (q,X)=\frac{4\pi}{\hbar^2}\sgn(q_0)\delta(q^2)\Big(
q^{\mu}q^{\nu}{f}_{\rm V}^{\gamma}
+ 2 \hbar q^{(\mu}S_{\gamma}^{\nu)\rho}\partial_{\rho}{f}_{\rm A}^{\gamma}
\Big)\,.
\end{eqnarray}
In the case of our counting scheme in which ${f}_{\rm A}^{\gamma} = \mathcal{O}(\hbar)$, 
the quantum correction will be at $\mathcal{O}(\hbar^2)$ and hence can be neglected.

\section{Derivation of the kinetic theory for photons with collisions}
\label{app_photon_KT}
To include the collisions, we follow the standard approach in Refs.~\cite{Blaizot:1999xk,Blaizot:2001nr,Hidaka:2016yjf}
with the Dyson-Schwinger equation 
\begin{eqnarray}
G^{-1}_{\mu\nu}=G_{(0)\mu\nu}^{-1}+\selfEnergy_{\mu\nu},
\end{eqnarray}
where $G_{\mu\nu}$ and $G_{(0)\mu\nu}$ represent the full and collisionless Feynman propagators for photons (right-handed plus left-handed) and $\Sigma_{\mu\nu}$ corresponds to the self-energy. 
In the iterating form, we have
\begin{eqnarray}\label{DS_eq_1}
G^{\mu\nu}(x,y)+\int {\rm d}^4 w{\rm d}^4 z G^{\mu\rho}_{(0)}(x,w)\selfEnergy_{\rho\sigma}(w,z)G^{\sigma\nu}(z,y)=G^{\mu\nu}_{(0)}(x,y)\,.
\end{eqnarray}
Here, $G^{\mu\nu}_{(0)}$ satisfies
\begin{eqnarray}
\Box^{\mu}_{x\,\rho} G^{\rho\nu}_{(0)}(x,y)={\rm i} \epsilon^{\mu\nu}(x)\delta^{(4)}_{\rm c}(x-y),
\end{eqnarray}
where $\Box^{\mu}_{x\,\rho}\equiv \big(-\delta^{\mu}_{~\rho}\partial^2+\partial_{\rho}\partial^{\mu}\big)_{x}$, 
$\epsilon^{\mu\nu}(x)$ represents the polarization tensor with a gauge dependence, and $\delta_{\rm c}^{(4)}(x-y)$ corresponds to the delta function on the Schwinger-Keldysh (SK) contour. 
Acting $\Box^{\mu}_{x\,\rho}$ on Eq.~(\ref{DS_eq_1}), we have
\begin{eqnarray}\label{DS_eq_2}
\Box^{\mu}_{x\,\rho} G^{\rho\nu}(x,y)+{\rm i}\int {\rm d}^4 z \epsilon(x)^{\mu\rho}\selfEnergy_{\rho\sigma}(x,z)G^{\sigma\nu}(z,y)={\rm i}\epsilon^{\mu\nu}(x)\delta^{(4)}_{\rm c}(x-y)\,.
\end{eqnarray}

Now, we focus on the parts of Eq.~(\ref{DS_eq_2}) involving the lesser propagator $G^{<\mu\nu}(x,y)$ for $x\neq y$. 
By utilizing the definitions 
\begin{eqnarray}\nonumber
G^{\mu\nu}(x,y)&\equiv&\theta_{\rm c}(x_0-y_0)G^{>\mu\nu}(x,y)+\theta_{\rm c}(y_0-x_0)G^{<\mu\nu}(x,y),
\\
\selfEnergy^{\mu\nu}(x,z)&\equiv&\theta_{\rm c}(x_0-z_0)\selfEnergy^{>\mu\nu}(x,z)+\theta_{\rm c}(z_0-x_0)\selfEnergy^{<\mu\nu}(x,z)-{\rm i}\selfEnergy^{(\delta)\mu\nu}(x)\delta^{(4)}_{\rm c}(x-z),
\end{eqnarray}  
and considering the SK contour, we obtain
\begin{eqnarray}\nonumber
0&=&\Box^{\mu}_{x\,\rho}G^{<\rho\nu}(x,y)+{\rm i}\int^{x_0}_{t_0}{\rm d}^4 z\epsilon^{\mu\rho}(x)\selfEnergy^>_{\rho\sigma}(x,z)G^{<\sigma\nu}(z,y)
+{\rm i}\int^{y_0}_{x_0}{\rm d}^4 z\epsilon^{\mu\rho}(x)\selfEnergy^<_{\rho\sigma}(x,z)G^{<\sigma\nu}(z,y)
\\
&&+{\rm i}\int^{t_0-{\rm i}\beta}_{y_0}{\rm d}^4 z\epsilon^{\mu\rho}(x)\selfEnergy^<_{\rho\sigma}(x,z)G^{>\sigma\nu}(z,y)
{+}\epsilon^{\mu\rho}(x)\selfEnergy^{(\delta)}_{\rho\sigma}(x)G^{<\rho\nu}(x,y),
\end{eqnarray}
which can be further written as
\begin{eqnarray}\nonumber\label{SK_eq}
0&=&\Box^{\mu}_{x\,\rho}G^{<\rho\nu}(x,y)+{\rm i}\int^{x_0}_{t_0}{\rm d}^4 z\epsilon^{\mu\rho}(x)\left(\selfEnergy^>(x,z)-\selfEnergy^<(x,z)\right)_{\rho\sigma}G^{<\sigma\nu}(z,y)
-{\rm i}\int^{y_0}_{t_0}{\rm d}^4 z\epsilon^{\mu\rho}(x)\selfEnergy^<_{\rho\sigma}(x,z)
\\\nonumber
&&\times\left(G^>(z,y)-G^<(z,y)\right)^{\sigma\nu}
+{\rm i}\int^{t_0-{\rm i}\beta}_{t_0}{\rm d}^4 z\epsilon^{\mu\rho}(x)\selfEnergy^<_{\rho\sigma}(x,z)G^{>\sigma\nu}(z,y)
+\epsilon^{\mu\rho}(x)\selfEnergy^{(\delta)}_{\rho\sigma}(x)G^{<\sigma\nu}(x,y),
\\
\end{eqnarray}
where $t_0$ and $t_0-{\rm i}\beta$ correspond to the initial time and final time in the SK contour. 

$G^{<\mu\nu}(x,y)$ should be independent of $t_0$ at $t_0\rightarrow -\infty$, so that we drop $t_0$ dependence by taking $t_0\rightarrow -\infty$. The last integrals along the imaginary-time axis in (\ref{SK_eq}) thus vanish. For the remaining integrals with respect to real time, we may rewrite the upper bounds of the integral as $\infty$ since the integrations from $x_0$ to $\infty$ (or from $y_0$ to $\infty$) do not contribute. Now, by introducing the retarded and advanced propagators
\begin{eqnarray}\nonumber\label{R_A}
G^{\mu\nu}_{\rm ret}(x,y)&\equiv& {\rm i}\theta(x_0-y_0)\left[G^>(x,y)-G^<(x,y)\right]^{\mu\nu},
\\
G^{\mu\nu}_{\rm adv}(x,y)&\equiv& -{\rm i}\theta(y_0-x_0)\left[G^>(x,y)-G^<(x,y)\right]^{\mu\nu},
\end{eqnarray}  
and similar definitions for the self-energy, Eq.~(\ref{SK_eq}) becomes
\begin{eqnarray}\label{SK_eq_2}
&&\big[\Box^{\mu}_{x\,\rho}+ \epsilon^{\mu\sigma}(x)\selfEnergy^{(\delta)}_{\sigma\rho}(x)\big]G^{<\rho\nu}(x,y)\notag\\
&&\quad=-\int^{\infty}_{-\infty} {\rm d}^4z\left[\epsilon^{\mu\rho}(x)\selfEnergy^{\rm ret}_{\rho\sigma}(x,z)G^{<\sigma\nu}(z,y)+\epsilon^{\mu\rho}(x)\selfEnergy^<_{\rho\sigma}(x,z)G^{\sigma\nu}_{\rm adv}(z,y)\right].
\end{eqnarray}
Introducing $G^{+\mu\nu}(x,y)= [G_{\rm ret}(x,y)+G_{\rm adv}(x,y) ]/2$, 
and $\selfEnergy^{+\mu\nu}(x,y)= [ \selfEnergy_{\rm ret}(x,y)+\selfEnergy_{\rm adv}(x,y) ]/2$,
we can express $G_{\rm ret/adv}^{\mu\nu}$ and  $\selfEnergy_{\rm ret/adv}^{\mu\nu}$ as
\begin{align}
G^{\mu\nu}_{\rm ret/adv}(x,y) &=G^{+\mu\nu}(x,y) \pm\frac{\ri}{2}\left[ G^{>\mu\nu}(x,y)  -G^{<\mu\nu}(x,y) \right]\,,  \\
\selfEnergy^{\mu\nu}_{\rm ret/adv}(x,y) &=\selfEnergy^{+\mu\nu}(x,y)\pm\frac{\ri}{2}\left[ \selfEnergy^{>\mu\nu}(x,y)  -\selfEnergy^{<\mu\nu}(x,y) \right]\,.
\end{align}
Using these expressions, we find  Eq.~\eqref{SK_eq_2} become
\begin{align}\label{SK_eq_3}
&\big[\Box^{\mu}_{x\,\rho}+\epsilon^{\mu\sigma}(x)\selfEnergy^{(\delta)}_{\sigma\rho}(x)\big]G^{<\rho\nu}(x,y)\notag\\
&\quad=\frac{\ri}{2} \int^{\infty}_{-\infty} {\rm d}^4z\epsilon^{\mu\rho}(x)
\bigl[\selfEnergy^<_{\rho\sigma}(x,z)G^{>\sigma\nu}(z,y) - \selfEnergy^{>}_{\rho\sigma}(x,z)G^{<\sigma\nu}(z,y) \bigr]
\notag\\
&\quad
-\int^{\infty}_{-\infty} {\rm d}^4z\epsilon^{\mu\rho}(x)\bigl[\selfEnergy^+_{\rho\sigma}(x,z)G^{<\sigma\nu}(z,y)+\selfEnergy^<_{\rho\sigma}(x,z)G^{+\sigma\nu}(z,y)
\bigr]\,.
\end{align}

We may subsequently implement the Wigner transformation and work in the Coulomb gauge. Note that 
\begin{eqnarray}\label{epsilon_transf}\nonumber
&&\epsilon^{\mu\nu}(x)\rightarrow {\cal P}^{\mu\nu}(q)= P^{\mu\nu}(q)-\frac{{\rm i}\hbar}{2}\delta P^{\mu\nu}(q) +\mathcal{O}(\hbar^2)\,,
\\
&&\delta P^{\mu\nu}(q)=\frac{1}{|\bm q|^2}\left(q^{(\mu}_{\perp}\partial^{\nu)}_{\perp}+q^{(\mu}_{\perp}\hat{q}^{\nu)}_{\perp}\hat{q}_{\perp}\cdot\partial_{\perp}\right)\,,
\end{eqnarray}
following the Wigner transformation in our setup, where $\delta P^{\mu\nu}(q)=-\big[\partial_{q_{\alpha}}P^{\mu\nu}(q)\big]\partial_{\alpha}$ is determined by the structure of $P^{\mu\nu}(q)$. One may check the $\hbar$ term in ${\cal P}^{\mu\nu}(q)$ is essential to satisfy the gauge constraint.   
To be more precise, by taking $\partial_{\perp \mu}^x$ of Eq.~(\ref{DS_eq_2}), the gauge constraint yields $\partial_{\perp \mu}^x \big[\epsilon^{\mu\rho}(x)\tilde{\selfEnergy}_{\rho}^{\,\,\nu}(x,y)\big]=0$ with $\tilde{\selfEnergy}_{\rho}^{\,\,\nu}(x,y)=\int {\rm d}^4 z \selfEnergy_{\rho\sigma}(x,z)G^{\sigma\nu}(z,y)$. Given $\epsilon^{\mu\rho}(x)$ is an operator, we find $\big(q_{\perp\mu}+{\rm i}\hbar\partial_{\perp\mu}/2\big){\cal P}^{\mu\rho}(q)\tilde{\selfEnergy}_{\rho}^{\,\,\nu}(q,X)=0$ with $\tilde{\selfEnergy}_{\rho}^{\,\,\nu}(q,X)$ being the dual function of $\tilde{\selfEnergy}_{\rho}^{\,\,\nu}(x,y)$ after the Wigner transformation. One can directly show $\big(q_{\perp\mu}+{\rm i}\hbar\partial_{\perp\mu}/2\big){\cal P}^{\mu\rho}(q)=0$ and thus the gauge constraint is always satisfied.
We assume a weak-coupling theory, so that we, hereafter, assign $\hbar$ to the self-energy. 
The Wigner transformation of (\ref{SK_eq_3})  takes the form
\begin{align}\label{SK_eq_4}
&\Big[\Box^{\mu}_{x\,\rho}+\frac{1}{\hbar}{\cal P}^{\mu\sigma}(\selfEnergy^{(\delta)}_{\sigma\rho}+\selfEnergy^+_{\rho\sigma})\star\Big] G^{<\rho\nu}
+\frac{1}{\hbar}{\cal P}^{\mu\rho}\selfEnergy^<_{\rho\sigma}\star G^{+\sigma\nu}=\frac{\ri}{2\hbar}{\cal P}^{\mu\rho}
\bigl(\selfEnergy^<_{\rho\sigma}\star G^{>\sigma\nu} - \selfEnergy^{>}_{\rho\sigma}\star G^{<\sigma\nu} \bigr)\,.
\end{align}
where $G^{\lessgtr, +}_{\mu\nu}$ and $\Sigma^{\lessgtr, +}_{\mu\nu}$ are now functions of $X$ and $q$, and
\begin{eqnarray}
\tilde{\Box}^{\mu}_{\,\,\rho}(X,q)\equiv \frac{\eta^{\mu}_{\mbox{ }\rho}}{\hbar^2}\Big(q^2-\frac{\hbar^2}{4}\partial^2+{\rm i}\hbar q\cdot\partial\Big)
-\frac{1}{\hbar^2} \Big(q^{\mu}+\frac{{\rm i}\hbar}{2}\partial^{\mu}\Big) \Big(q_{\rho}+\frac{{\rm i}\hbar}{2}\partial_{\rho}\Big)\,.
\end{eqnarray}
We also introduced the Moyal product such that 
\begin{eqnarray}
 A(q,X)\star B(q,X) = \int {\rm d}^4Y {\rm e}^{{\rm i}q\cdot Y/\hbar} \int \mathrm{d}^4z A(X+Y/2,z)B(z,X-Y/2)\,,
\end{eqnarray}
which can be expanded in terms of $\hbar $ as
\begin{align}
A(q,X)\star B(q,X)= A(q,X)B(q,X)+\frac{{\rm i}\hbar}{2}A(q,X) * B(q,X) + \mathcal{O}(\hbar^2)\,,
\end{align}
where $A* B \equiv (\partial_{q_{\alpha}} A)(\partial_{X^\alpha}B )-(\partial_{X^\alpha} A)(\partial_{q_{\alpha}}B)$ is a shorthand notation of the Poisson bracket. 

The term $\selfEnergy^{(\delta)}_{\sigma\rho}+\selfEnergy^+_{\rho\sigma}$ in Eq.~\eqref{SK_eq_4} gives the self-energy correction, which we drop since we are interested in the collisional effects. 
We also drop the term proportional to ${\cal P}^{\mu\rho}\selfEnergy^<_{\rho\sigma}\star G^{+\sigma\nu}$ because its contribution is negligible compared with $G^<$ for on-shell photons.
Eventually, Eq.~(\ref{SK_eq_4}) reduces to
\begin{align}\label{SK_eq_5}
&\Big(q^2-\frac{\hbar^2}{4}\partial^2+\ri\hbar q\cdot \partial\Big) G^{<\mu\nu}
=\frac{\ri\hbar}{2}{\cal P}^{\mu\rho}
\bigl(\selfEnergy^<_{\rho\sigma}\star G^{>\sigma\nu} - \selfEnergy^{>}_{\rho\sigma}\star G^{<\sigma\nu} \bigr).
\end{align}

We may now decompose Eq.~(\ref{SK_eq_5}) into the real and imaginary parts. By decomposing the greater/lesser self-energies into the real and imaginary parts as 
$\selfEnergy_{\rho\sigma}=(\selfEnergy_{\text{Re}})_{\rho\sigma}+{\rm i}(\selfEnergy_{\text{Im}})_{\rho\sigma}$, we find
\begin{align}
&\Big(q^2-\frac{\hbar^2}{4}\partial^2+\ri\hbar q\cdot \partial\Big)( G_{\rm S}^{<\mu\nu}+ \ri G_{\rm A}^{<\mu\nu})\notag\\
&=-\frac{1}{2}P^{\mu\rho}
\left[
\ri\widehat{(\selfEnergy_{\mathrm{Re}} G_{\rm S}})_{\rho}^{~\nu} -\ri\widehat{(\selfEnergy_{\mathrm{Im}} G_{\rm A}})_{\rho}^{~\nu} 
-\widehat{(\selfEnergy_{\mathrm{Im}} G_{\rm S}})_{\rho}^{~\nu} 
-\widehat{(\selfEnergy_{\mathrm{Re}} G_{\rm A}})_{\rho}^{~\nu} 
\right]\notag\\
&-\frac{\hbar}{4}\delta P^{\mu\rho}
\left[  (\widehat{\selfEnergy_{\mathrm{Re}} G_{\rm S}})_\rho^{~\nu}-(\widehat{\selfEnergy_{\mathrm{Im}} G_{\rm A}})_\rho^{~\nu}   
+\ri(\widehat{\selfEnergy_{\mathrm{Im}} G_{\rm S}})_\rho^{~\nu}  +\ri(\widehat{\selfEnergy_{\mathrm{Re}} G_{\rm A}})_\rho^{~\nu}  
 \right]\notag\\
&+\frac{\hbar}{4} P^{\mu\rho}
\left[  (\widehat{\selfEnergy_{\mathrm{Re}} ^{>}* G_{\rm S}^{<}})_\rho^{~\nu}-(\widehat{\selfEnergy_{\mathrm{Im}} ^{>}* G_{\rm A}^{<}})_\rho^{~\nu} 
+\ri(\widehat{\selfEnergy_{\mathrm{Im}} ^{>}* G_{\rm S}^{<}})_\rho^{~\nu} 
+\ri(\widehat{\selfEnergy_{\mathrm{Re}} ^{>}* G_{\rm A}^{<}})_\rho^{~\nu} 
 \right] +\mathcal{O}(\hbar^3)\,,
\end{align}
where we introduced shorthand notations, $(AB)_{\rho}^{\, \ \nu}\equiv A_{\rho\sigma}B^{\sigma\nu}$ and $(\widehat{A B})_{\rho}^{\ \,\nu}\equiv A^{>}_{\rho\sigma}B^{<\sigma\nu}-A^{<}_{\rho\sigma}B^{>\sigma\nu}$. By further separating the symmetric and anti-symmetric parts, we derive the constraint and kinetic equations up to $\mathcal{O}(\hbar)$, 
\begin{align}\label{cons_eq_full_1}
q^2G^{<\mu\nu}_{\rm S}&=\frac{\hbar}{4} P^{(\mu\rho}
\left[(\widehat{\selfEnergy_{\text{Re}} G_{\rm A}})_{\rho}^{\,\,\nu)}+(\widehat{\selfEnergy_{\text{Im}} G_{\rm S}})_{\rho}^{\,\,\nu)}
\right]\,,
\\
q^2G^{<\mu\nu}_{\rm A}&=-\frac{\hbar}{4} P^{[\mu\rho}
\left[(\widehat{\selfEnergy_{\text{Re}} G_{\rm S}})_{\rho}^{\,\,\nu]}-(\widehat{\selfEnergy_{\text{Im}} G_{\rm A}})_{\rho}^{\,\,\nu]}
\right]\,,
\\
\nonumber\label{kin_eq_full_S}
q\cdot\partial G^{<\mu\nu}_{\rm S}
&=-\frac{1}{4} P^{(\mu\rho}
\left[(\widehat{\selfEnergy_{\text{Re}} G_{\rm S}})_{\rho}^{\,\,\nu)}-(\widehat{\selfEnergy_{\text{Im}} G_{\rm A}})_{\rho}^{\,\,\nu)}
\right]
-{\frac{\hbar}{8}\delta P^{(\mu\rho}(q)}
\left[(\widehat{\selfEnergy_{\text{Re}} G_{\rm A}})_{\rho}^{\,\,\nu)}+(\widehat{\selfEnergy_{\text{Im}} G_{\rm S}})_{\rho}^{\,\,\nu)}\right]
\\
&\quad+\frac{\hbar}{8}P^{(\mu\rho}\left[(\widehat{\selfEnergy_{\text{Re}}* G_{\rm A}})_{\rho}^{\,\,\nu)}+(\widehat{\selfEnergy_{\text{Im}}* G_{\rm S}})_{\rho}^{\,\,\nu)}
\right]\,,\\
q\cdot\partial G^{<\mu\nu}_{\rm A}
&=-\frac{1}{4} P^{([\mu\rho}\left[(\widehat{\selfEnergy_{\text{Re}} G_{\rm A}})_{\rho}^{\,\,\nu]}+(\widehat{\selfEnergy_{\text{Im}} G_{\rm S}})_{\rho}^{\,\,\nu]}
\right]
+{\frac{\hbar}{8}\delta P^{[\mu\rho}(q)}\left[(\widehat{\selfEnergy_{\text{Re}} G_{\rm S}})_{\rho}^{\,\,\nu]}-(\widehat{\selfEnergy_{\text{Im}} G_{\rm A}})_{\rho}^{\,\,\nu]}
\right]\notag\\
&\quad-\frac{\hbar}{8}P^{([\mu\rho}\left[(\widehat{\selfEnergy_{\text{Re}}* G_{\rm S}})_{\rho}^{\,\,\nu]}-(\widehat{\selfEnergy_{\text{Im}}* G_{\rm A}})_{\rho}^{\,\,\nu]}
\right]\,.
\end{align}

The $\hbar$ terms in the kinetic theory are essential to satisfy the gauge constraint. For example, we may show that Eq.~(\ref{SK_eq_5}) as a master equation of the kinetic theory satisfies the gauge constraint. 
Contracting the left/right-hand sides of the kinetic equation~(\ref{SK_eq_5}) with $q_{\perp\mu}$, we have 
\begin{eqnarray}\nonumber
\text{LHS}&=& q_{\perp\mu}\Big(q^2-\frac{\hbar^2}{4}\partial^2+{\rm i}\hbar q\cdot\partial\Big)G^{<\mu\nu}
\\\nonumber
&=&-\frac{{\rm i}\hbar}{2}\partial_{\perp\mu}\Big(q^2-\frac{\hbar^2}{4}\partial^2+{\rm i}\hbar q\cdot\partial\Big) G^{<\mu\nu}
\\
&=&-\frac{\hbar^2}{4}\partial_{\perp\mu}P^{\mu\rho}\big(\widehat{\selfEnergy G})^{\,\,\nu}_{\rho}+\mathcal{O}(\hbar^3),
\end{eqnarray}
and
\begin{eqnarray}\nonumber
\text{RHS}&=& -\frac{\hbar}{2}q_{\perp\mu}\Big[{\cal P}^{\mu\rho}\ri(\widehat{\selfEnergy G})_{\rho}^{\,\,\nu}
+\frac{{\rm i}\hbar}{2}{\rm i}P^{\mu\rho}(\widehat{\selfEnergy * G})_{\rho}^{\,\,\nu}
\Big]
\\\nonumber
&=&\frac{\hbar^2}{4}\big(\partial_{\perp}^{\rho}+\hat{q}^{\rho}_{\perp}\hat{q}_{\perp}\cdot\partial_{\perp}\big)(\widehat{\selfEnergy G})_{\rho}^{\,\,\nu}+\mathcal{O}(\hbar^3)
\\
&=&-\frac{\hbar^2}{4}P^{\mu\rho}\partial_{\perp\mu}\big(\widehat{\selfEnergy G})^{\,\,\nu}_{\rho}+\mathcal{O}(\hbar^3),
\end{eqnarray}
respectively. It turns out that the LHS is equal to the RHS up to $\mathcal{O}(\hbar^2)$. The proof above also justifies the inclusion of necessary $\hbar$ correction in Eq.~(\ref{epsilon_transf}).

\bibliography{WF_QKT_journal.bbl}
\end{document}